\documentstyle[epsf,referee]{mn}

\newcommand{\rmn}[1] {{\rm #1}}

\author[L.V.E. Koopmans et al.]{L.V.E. Koopmans$^1$, A.G. de
  Bruyn$^{1,2}$ \& N. Jackson$^3$\\ $^1$Kapteyn Astronomical
  Institute, P.O. Box 800, 9700 AV Groningen, The Netherlands\\ 
  $^2$NFRA, P.O. Box 2, 7990 AA Dwingeloo, The Netherlands\\ 
  $^3$NRAL Jodrell Bank, University of Manchester, Macclesfield,
  Cheshire SK11 9DL, UK} \date{Accepted ; Received } \pubyear{1997}
\title{The edge-on spiral gravitational lens B1600+434}

\begin{document}
\maketitle

\begin{abstract}
  We present new observations of the gravitational lens (GL) system
B1600+434, strongly suggesting that the lens is an edge-on spiral
galaxy. These observations are used to constrain the mass model of the
system, in particular the oblateness and velocity dispersion of the
dark matter halo around the lensing galaxy. From an analytical model
we find a lower limit on the halo oblateness
$q_\rmn{halo}=(c/a)_{\rho}\ga0.4$; more detailed numerical models give
a lower limit $q_\rmn{h}\ga0.5$. We determine an average halo velocity
dispersion of $\sigma_{\rm halo}=190\pm 15$ km/s over all non-singular
isothermal elliptical (NIE) halo models.  Constraining the models to
larger and more massive disks, decreases this average by only 10
km/s. A lower limit of $\sigma_{\rm halo}\ga 150$ km/s is found, even
for disk masses larger then the mass inside the Einstein radius. This
lower limit indicates the need for a massive dark matter halo,
contributing at least half of the mass inside the Einstein radius.
Time delay calculations give $(54\pm3)/h_{50}$ days for the NIE halo
model and $(70\pm4)/h_{50}$ days for the modified Hubble profile (MHP)
halo model.  Although the time delay for both NIE and MHP halo models
is well constrained on our parameter grid, it strongly depends on the
halo surface density profile.  We furthermore find that the presence
of a flat luminous mass distribution can severely alter the
statistical properties of the lens.

\end{abstract}


\section{Introduction}
Over the last two decades, gravitational lensing has proved a very
effective tool in constraining the shape of the mass distribution
responsible for weak and strong lensing.  Moreover, a time delay
between two images can be used to determine the Hubble parameter,
given an appropriate mass model (Refsdal 1964; 1966). In this paper we
focus on the oblateness and velocity dispersion of the dark matter
halo around the edge-on spiral galaxy lens in the GL system B1600+434
and calculate the expected time delays for this system for two
different halo mass models.

Different techniques have been used to determine the oblateness
$q_\rmn{halo}=(c/a)_\rmn{\rho}$ of the dark matter halo around spiral
galaxies.  Most of these indicate an oblateness $\ga 0.4$
(dissipationless N-body calculations, stellar dynamics in the solar
neighborhood and polar ring galaxies (e.g.  Rix 1995; Sackett 1995)).
On the other hand, flaring of the gas layer and the HI velocity
dispersion in NGC 4244 seem to indicate a more oblate dark matter halo
with $q_\rmn{halo}=0.1-0.5$ for that galaxy (Olling, 1996). This could
support the idea that dark matter is mostly baryonic and perhaps
consists of molecular hydrogen (Pfenniger, Combes \& Martinet 1994;
Pfenniger \& Combes 1994).  Obviously one would like to put stronger
constraints on the oblateness of dark matter halos around spiral
galaxies, less dependent of the chosen mass model, assumptions about
the type of dark matter or its dynamical state.  This opportunity is
offered by the GL system B1600+434.

The lensing galaxy of B1600+434 is an edge-on spiral galaxy between
two lensed QSO images and therefore presents the opportunity to
determine the oblateness of the halo $(c/a)_\rho$ by means of lensing.
This method is different from and independent of methods used
previously, in the sense that it is not dependent on the type of
lensing matter (e.g. baryonic or non-baryonic) or its dynamical state.

In section 2 we describe new observations of B1600+434; in section 3
the basic lensing theory is summarized; in section 4 the mass models
used to describe the lensing galaxy, its dark matter halo and the
companion galaxy are presented; in section 5 we describe the parameter
space of the fixed parameters in the mass models; and in section 6 we
present our results and analyses. The main results are summarized in
section 7.

\section{Observations}
The double QSO lens system B1600+434 was discovered by Jackson et al.
(1995) in the Cosmic Lens All Sky Survey (CLASS; Myers et
al. 1997). First we briefly summarize the observational status of this
system and then present new radio and optical observations. 

\paragraph*{Observational status}

   B1600+434 was discovered as a compact flat spectrum radio double
with a separation of 1.39$\pm$0.01$''$. VLA 8.4 GHz observations show
no structure due to extended emission (e.g.  lobes) around the compact
flat spectrum radio core and no sign of a central third image above
the noise level is found (Jackson et al.  1995).  On February 24 1996,
B1600+434 was observed with the VLBA at 6 and 18 cm to search for
possible substructure in the radio maps, which can further constrain a
mass model for the lensing galaxy. The VLBA 18 cm maps of QSO images A
and B do not show signs of substructure, but the 6 cm images show
possible substructure 1--2 mas from the brightest component (Jackson,
private communications).  The redshift of the lensed source was
measured to be 1.61 with the WHT (Jackson et al. 1995). Subsequently,
more accurate redshifts of the source and the lens galaxy were
determined with the Keck telescope (Fassnacht et al.  1998).  The
redshift of G1 was determined at 0.415 and the redshift of the source
1.59, consistent with the source redshift found previously.  We will
use the Keck redshifts, because of their superior S/N.

\paragraph*{HST observations}

On November 18 1995, a 700 seconds WFPC2 HST I band and an 800 seconds
V band exposure were obtained of B1600+434. The I band image clearly
shows the two QSO images A and B, the lensing galaxy (G1) between the
QSO images and the companion galaxy (G2) south-east of G1 (Fig. 1).
The luminous component of G1 appears to be a flat edge-on system that
exhibits a prominent dust lane. Both the dust lane and flat luminous
mass distribution indicate that G1 is most likely a nearly edge-on
spiral galaxy. The V band exposure shows both images A and B and
galaxy G2, but not the lens galaxy G1. 

We performed I and V band photometry on galaxies G1 and G2, on the
small ($\sim3''$) moderately inclined disk galaxy $\sim10''$ west of
G1 (G3; see Jaunsen and Hjorth (1997) (JH97)) and on the QSO images A
and B. We applied corrections for the gain of the different chips, the
transfer efficiency and the long exposure time, amounting to a total
correction in both V and I of approximately $-0.08$ magnitude.  We
also determined the magnitudes of two field stars S1 and S2 (stars 1
and 2 in JH97).  The photometric results are listed in Table
1. Comparing our HST photometry with the ground based photometry of
JH97, we find that the magnitudes of all three galaxies and both stars
do not differ by more than $\sim0.1$ magnitude in both V and I.
However, over the same period (July to November 1995) the QSO images A
and B have dimmed significantly. Image A by some 0.5-0.6 magnitudes in
V and 0.5 in I and image B by some 0.2 magnitudes in both V and I
bands. The decrease in brightness of image A is very significant and
indicates strong optical variability over the order of months.  The
smaller decrease in brightness of image B can be explained by a time
delay between the two images, which is also in the order of months
(see section 6), but subtracting the emission of G1 could also have
caused a systematic error in the brightness determination of image B.
This effect will be stronger for less resolved ground based
observations.

The companion galaxy G2 south-east of G1 appears to be a nearly
face-on luminous barred spiral galaxy (Fig. 1). The bar-like structure
can also be seen in the deconvolved NOT R band image (Fig. 2).  From
Fukugita, Shimasaku \& Ichikawa (1995) we find that the HST
F555W-F814W color of $1.40\pm0.15$ could either indicate an early type
spiral galaxy at relative low redshift ($\sim 0.2$) or a somewhat
later type spiral galaxy at higher redshift ($\sim 0.5$). Judging from
the bar-like structure G2 does not appear to an E/S0 galaxy, as
suggested by JH97. As the photometric redshift of G2 is rather
uncertain, we assume it to be the same as for G1 (0.415) as most
reasonable first estimate.  In section 6 we will describe the
dependence of our results on this assumption.  The small galaxy G3 is
a moderately inclined (spiral) galaxy some $10''$ west of
B1600+434. Because this galaxy is much smaller and $\sim$2 magnitudes
fainter than G2 ($\sim 6$ times less massive for the same M/L ratio
and redshift), we do not incorporate this galaxy in the mass models
(both the convergence and shear of G3 will be $\sim10$ times smaller
than that of G2 at images A and B). 

\paragraph*{Nordic Optical Telescope observations}

Exposures in B and R band (both 600 seconds) were taken July 30 1995
with the BroCam 1 camera (TEK1024 CCD) on the Nordic Optical Telescope
(NOT). They show the lensing galaxy G1, the two QSO images and galaxy
G2 (Fig. 2).  Comparing the two QSO images in B and R, we clearly see
the effect of extinction on image B, which almost passes through the
dust lane of G1 (Fig. 1).  To enhance the resolution, we deconvolved
the R band image with the maximum entropy method `{\tt mem}' in the
IRAF package STSDAS.  The result clearly shows the extent of G1
($\sim7''$) and G2 ($\sim5''$). There seems to be no clear evidence in
the NOT and HST images to support the presence of a prominent massive
bulge component in galaxy G1. 

\paragraph*{MERLIN observations}

On March 14 1995, a MERLIN 5 GHz observation was made of B1600+434
(Fig. 3). The map shows two compact ($\sim 50$ mas) radio components,
with no visible sign of extended emission above the noise level. The
flux densities of both components are given in Table 1. The flux ratio
between components A and B is $\sim1.2$, comparable to the flux ratio
from the VLA observations of Jackson et al.  (1995). This consistency
in flux ratio is either a coincidence (although VLA observation in
August 1995 also indicate a flux ratio of $\sim1.2$ at 8.4 GHz, see
Table 1) or the typical time scale of variability in the radio is much
larger than the time delay, resulting only in slight variations in the
radio flux ratio. 

\paragraph*{Variability of the source}

VLA 8.4 GHz observations of B1600+434 at two epochs indicate that the
lensed source is variable by at least a factor of two over a period of
one year (Table 1).  Moreover, observations over a period of three
months (April to July 1996) at 21 cm continuum with the WSRT (Fig. 4)
indicate variability on time scales of the order of the expected time
delay (section 6.3). It appears there is a steady decrease of flux
density over this period, totaling $\sim 10\%$ over three months. 

This variability does not appear to strongly affect the flux ratio in
the radio, which stayed between 1.2-1.3 at three epochs over a period
of some two years.  A ratio between the QSO images was found to be
1.38$\pm$0.05 in R (Jackson et al.  1995). Our new HST I band
magnitudes give a ratio of $1.2\pm0.2$. The observations of JH97 give
a ratio of 1.6 in I (epoch 1). Although they could have underestimated
the I magnitude of image B as a result of subtracting the lensing
galaxy, it can also indicate a much stronger and perhaps more rapid
variability in the optical.  Furthermore, optical ratios will be
effected by dust extinction and therefore give more or less upper
limits, even if corrected for time delay.  Overall our radio flux
ratios appears to stay consistently around 1.2, whereas the optical
ratios are slighly higher.  We adopt a ratio $r_\rmn{AB}=1.25 \pm
0.10$.

\section{Lensing theory}
In describing basic lensing theory we will follow the definitions and
notations as in Schneider, Ehlers \& Falco (1992).\\

\smallskip \noindent Given a length scale $\xi_0$ in the lens plane
and the corresponding length scale $\eta_0=\xi_0 D_s/D_d$ in the
source plane we can define the dimensionless vectors

\begin{equation}
 {\bf x}={\bf \xi}/\xi_0~~\mbox{and}~~{\bf y}={\bf \eta}/{\eta_0}.
\end{equation}
We can then define the dimensionless surface mass density
\begin{equation}
   \kappa({\bf x})=\Sigma(\xi_0 {\bf x})/\Sigma_{cr},
\end{equation}
where the critical surface mass density is given by
\begin{equation}
   \Sigma_{cr}=\frac{c^2 D_s}{4 \pi G D_d D_{ds}}.
\end{equation}
The lens equation 
\begin{equation}
   {\bf \eta}=\frac{D_s}{D_d}{\bf \xi}-D_{ds}\hat{\bf \alpha}({\bf \xi})
\end{equation}
then becomes
\begin{equation}
   {\bf y}={\bf x}-{\bf \alpha}({\bf x}),
\end{equation}
with
\begin{equation}
   {\bf \alpha}({\bf x})=\frac{1}{\pi}\int_{R^2}d^2{x'} 
                   \kappa({\bf x}')\frac{{\bf x}-{\bf x}'}{|{\bf x}-
                   {\bf x}'|^2}=\frac{D_s D_{ds}}{\xi_0 D_s} 
                   \hat{\bf \alpha}(\xi_0 {\bf x}).
\end{equation}
We can then define the deflection potential 
\begin{equation}
   \psi({\bf x})=\frac{1}{\pi}\int_{R^2} d^2 x' \kappa({\bf x}')
                 \ln|{\bf x}-{\bf x}'|,
\end{equation}
in order to get 
\begin{equation}
   {\bf \alpha}=\nabla \psi.
\end{equation}
The dimensionless lens equation then becomes
\begin{equation}
   {\bf y}=\nabla\left(\frac{1}{2}{\bf x}^2 - \psi({\bf x})\right),
\end{equation}
which can also be written, using
\begin{equation}
   \phi({\bf x},{\bf y})=\frac{1}{2}({\bf x}-{\bf y})^2 - \psi({\bf x})
\end{equation}
as
\begin{equation}
   \nabla\phi({\bf x}, {\bf y})=0.
\end{equation}
The image distortion from the source to the image plane can be
described by the Jacobian matrix
\begin{equation}
   A({\bf x})=\frac{\partial{\bf y}}{\partial{\bf x}}=
               \left[\begin{array}{cc}
                        1-\kappa-\gamma_1 & -\gamma_2 \\
                        -\gamma_2 & 1-\kappa+\gamma_1 \\
                     \end{array}\right]
\end{equation}
with $\kappa({\bf x})$ being the local surface mass density (convergence) 
and the shear components 
\begin{equation}
   \gamma_1=\frac{1}{2}(\psi_{11}-\psi_{22}),~~
                 \gamma_2=\psi_{12}=\psi_{21}.
\end{equation}
The magnification factor is then given by
\begin{equation}
   \mu({\bf x})=\frac{1}{\mbox{det}A({\bf x})}=\frac{1}{(1-\kappa)^2-
         \gamma^2}
\end{equation}
with
\begin{equation}
    \gamma^2=\gamma_1^2+\gamma_2^2.
\end{equation}
The time delay between to images at ${\bf x}^{(1)}$ and ${\bf x}^{(2)}$
is given by
\begin{equation}
   \Delta t=\frac{\xi_0^2}{c}\frac{D_s}{D_d D_{ds}}(1+z_d)\left[
            \phi\left({\bf x}^{(1)},{\bf y}\right)-
            \phi\left({\bf x}^{(2)},{\bf y}\right)\right].
\end{equation}
Because $\Delta t \propto \frac{1}{\rmn{H}_0}$ we can use the the
above equation to determine the Hubble parameter $\rmn{H}_0$, given
the observed time delay between two lensed images and an appropriate
mass model for the lens (Refsdal 1964; 1966).

\section{Mass model}
To model B1600+434, we construct a mass distribution consisting of 4 components: 
the luminous disk and bulge of G1, the dark matter halo around G1 and the 
combined luminous+dark matter distribution of galaxy G2. We will describe
these components separately below. We use a Cartesian coordinate system and
define our x$_1$-axis to lie along the dust lane. The origin is defined 
on the point where x$_1$ and the line joining A and B intersect. 
The line joining images A and B makes an angle of $17^\circ \pm 2^\circ$ 
with the x$_2$ axis, consistent with what we find from our NOT images and 
the angle of $15^\circ \pm 3^\circ$ derived from JH97. We determine 
the image and galaxy positions with respect to this fixed coordinate system
(Table 2). Fitting ellipses to the bright inner part of G1 (masking images
A and B) in I band we find that the center of the brightness distribution 
of G1 is consistent with our defined origin (Fig. 1). This indicates that 
there is no significant angle between image A, the origin and image B.
Lower surface brightness contours indicate a slight shift in the center
of the ellipse center towards x$_1<0$, but smaller than the difference in lens
center between us and Maller, Flores \& Primack (1997), who used
the results of JH97 to model B1600+434. The center of G1 that we use in this paper
is also consistent with the position of the surface brightness of G1 in recent 
NICMOS H band observations by Jackson et al. (private communications). But much
deeper optical or near infrared observations are still necessary to accurately pin 
down the center of G1. 

In all calculations we will assume a smooth Friedmann-Robertson-Walker (FRW) 
universe with $\Omega=1$, $\Lambda=0$ and $h_{50}=1$ (H${}_0=50\cdot h_{50}$ km/s/Mpc), 
if not explicitly specified otherwise. 

\subsection{Disk and bulge and halo of G1}

The surface brightness distribution of most disk galaxies can be described by
an exponential profile (e.g. Mihalas \& Binney 1981). Assuming a constant
mass-to-light (M/L) ratio, the surface mass distribution of the disk 
of G1 becomes
\begin{equation} 
	\Sigma_\rmn{G1,disk}(x_1,x_2)=\Sigma_{\rm G1,disk}^0 
		e^{-\sqrt{(x_1^2+f_{\rm disk}^{-2} \cdot x_2^2)/h^2_{\rm disk}}},
\end{equation}
where $h_{\rmn{disk}}$ is the radial exponential scale length and 
$f_\rmn{disk}$ is the axis ratio of the disk surface brightness (mass)
distribution projected on the sky. Although, when seen edge-on, this exponential 
surface mass distribution is not completely valid anymore, we assume 
that this relation stays valid in first order at large inclinations.

Many bulges can be well described by a de Vaucouleurs surface 
brightness profile, i.e. R${}^{1/4}$ law (e.g. Mihalas \& Binney 1981). 
By also assuming that the M/L ratio is constant for the bulge, we find
\begin{equation} 
	\Sigma_\rmn{G1,bulge}(x_1,x_2)=\Sigma_\rmn{G1,bulge}^0
		 e^{-7.67 [{(x_1^2+f_{\rm bulge}^{-2} \cdot x_2^2)/r^2_{e}}]^{1/8}}.
\end{equation} 
From de Jong (1996) we find $r_e\approx \frac{1}{7}\cdot h_{\rm disk}$
and assume this relation to hold for the effective radius of the bulge
of G1. We assume $f_{\rm bulge}=0.6$, as found for NGC891 (Bottema,
van der Kruit \& Valentijn 1991), which looks morphologically quite
similar to G1.  The optical extent of NGC891 ($\sim$30-35 kpc) is
somewhat but not significantly smaller to that of G1 ($\sim$45 kpc,
$h_{50}=1$).  The disk mass of G1 is assumed 30 times more massive
than the bulge mass (e.g. NGC891) in all our mass models. Although we
have tried several models with free disk and bulge masses, they all
give results in contradiction to our observations (section 6.1).

Because little is known about the actual surface mass distribution of
the halo around disk galaxies, we model the halo around G1 with two
different surface mass distributions: the non-singular isothermal
ellipsoid model (NIE) and the modified Hubble profile (MHP).
Following Kormann, Schneider \& Bartelmann (1994) we find for the NIE
models

\begin{equation} 
	\Sigma_\rmn{G1, halo}(x_1,x_2)=\frac{\Sigma_\rmn{G1,halo}^0 }
		{\sqrt{{1+(f_\rmn{halo}^{2} x_1^2+  x_2^2)/r_{\rm c}^2}}}
\end{equation}
where $r_{\rm c}$ is the halo core radius and $f_\rmn{halo}$ the flattening. 
The velocity dispersion is defined as 
\begin{equation} 
	\sigma_\rmn{halo}^2=2 G \Sigma_\rmn{halo}^0 r_\rmn{c} \sqrt{f_\rmn{halo}}.
\end{equation}
For the MHP models we use 
\begin{equation}
	\Sigma_\rmn{G1, halo}(x_1,x_2)=\frac{\Sigma_\rmn{G1,halo}^0 }
		{{{1+(f_\rmn{halo}^{2} x_1^2+  x_2^2)/r_{\rm c}^2}}}.
\end{equation}
We align the major axes of the disk, bulge and halo of G1 along the
x$_1$-axis and center them on the origin. Because G1 is close to
edge-on, we assume that the axis ratio of the surface mass
distribution of the halo is very close the its oblateness, hence
$f_{\rm halo}=q_{\rm halo}=(c/a)_{\rho, \rm halo}$.  For inclinations
$\ga75^\circ$ and $q_{\rm halo}\ga0.5$ the difference between $f_{\rm
halo}$ and $q_{\rm halo}$ is $\la10\%$.

\subsection{The galaxy G2}

Because galaxy G2 appears to be a nearly face-on luminous disk galaxy,
we use a non-singular isothermal sphere (NIS) as surface mass model,
consistent with the assumption that disk galaxies have a flat rotation
curve, as observed in many luminous nearby disk galaxies (e.g. Begeman
1987; Broeils 1992).  The projected distance between G1 and G2 is
$\sim30$ kpc, a distance at which most of these galaxies still have
flat rotation curves and the dark matter halo dominates the surface
mass density (e.g. Begeman 1987; Broeils 1992). The oblateness
$q^{G2}=(c/a)_{\rho}^{G2}$ of the mass distribution of G2 doesn't
influence the radial profile of the surface mass distribution.  We
therefore assume that a NIS is a reasonable model to describe the
surface mass distribution of G2 in first order

\begin{equation} 
	\Sigma_\rmn{G2}(x_1,x_2)=\frac{\Sigma_\rmn{G2}^0}{\sqrt{1+(x_1^2+x_2^2)/r_\rmn{G2,c}^2}},
\end{equation}
where $r_{\rm G2,c}$ is the core radius of the mass distribution of G2. 
The velocity dispersion is given by
\begin{equation} 
	\sigma_{\rm G2}^2=2 G \Sigma_{\rm G2}^0 r_{\rm G2,c}.
\end{equation}
We assume a small core radius of 0.1 kpc. We center G2 on the 
position given in table 1 and assume the redshifts of G1 and G2 to be the
same, as explained in section 2.

\section{Parameter space}
Our knowledge of the galaxies G1 and G2 is rather limited by the low
signal-to-noise and/or low resolution of the optical images.  To asses
the reliability of the results that we obtain on the halo flattening,
halo velocity dispersion and time delay between the two lensed images,
we examine a large grid of parameters that describe the mass
distributions of G1 and G2. We calculate a grid of $\sim5000$ models
using the NIE halo model, each with different core radius of the halo
($r_\rmn{c}$), velocity dispersion of galaxy G2 ($\sigma_\rmn{G2}$),
radial exponential scale length of the disk of G1 ($h_\rmn{disk}$),
disk mass of G1 ($M{}_\rmn{disk}$) and stellar disk flattening of G1
($f_\rmn{disk}$) (see Table 3). We do the same for the MHP halo model.
Keeping these parameters fixed, we vary the velocity dispersion
$\sigma_{\rm halo}$ and flattening $f_{\rm halo}$ of the halo of G1 to
achieve the minimum value of $\chi^2$ ($\chi^2_\rmn{min}$), where we
define (Kayser 1990)

\begin{equation}
	\chi^2 \cdot \rmn{N}_\rmn{dof}=\frac{|{\bf y}({\bf x}_\rmn{A}) - 
	{\bf y}({\bf x}_\rmn{B})|^2}{\delta y^2} +
	   \frac{|r_\rmn{AB}-\frac{|{\bf J}_\rmn{B}|}{|{\bf J}_\rmn{A}|}|^2}
	{\delta r_\rmn{AB}^2}.
\end{equation}
Here ${\bf y}({\bf x})$ is the lens equation, ${\bf x}_\rmn{A/B}$ are
the positions of lensed images A and B, $r_\rmn{AB}$ is the flux
ratio, ${\bf J}_\rmn{A/B}$ are the Jacobians at ${\bf x}_\rmn{A/B}$
and
N$_\rmn{dof}=\rmn{N}_\rmn{data}-\rmn{N}_\rmn{pars}$\footnote{$\rmn{N}_\rmn{data}$
is the number of constraints from the observations and
$\rmn{N}_\rmn{pars}$ is the number of free parameters in the mass
model.} is the number of degrees of freedom. For the flux ratio we
adopt an error $\delta r_\rmn{AB}=$0.1, as explained in section
2. Furthermore $\delta y$ is the position error of the source in the
source plane.  We examined different descriptions for $\delta y$,
because the circular error regions around the lensed images A and B do
in fact not project back to circles on the lens plane, but project
back onto ellipses.  First we adopted a circular error region with
$\delta y=0.02''$, which roughly corresponds to an error of $0.05''$
in the image plane for typical magnifications of a few. This error is
comparable to the error of the image positions with respect to the
lens center.  A choise of the $\delta y=0.005''$ does not change our
results significantly.  When projecting the error regions around the
lens images back on the source plane, we find that they become two
orthogonal ellipses. This increases the allowed region somewhat inside
which the two lens images can be projected back on the source
plane. However, redoing a sizeable subsample of our models indicates
only a slight change in results, being on average only a shift of
10-20\% of the rms value. This shift is therefore not significant.

Our choise of error region ($0.05''$ around both images) allows for a
spread in image seperation, which is more than the observed $0.01''$
(Jackson et al. 1995). So we also looked at those solutions with an
image separation of $1.39\pm0.01''$ and again find no significant
change in our results compared with the other methods. We are
therefore quite confident that our results are not strongly dependend
on the choise of the allowed region ($\delta y$) in the source plane
(e.g. the topology of the $\chi^2$-space is quite robust as function
of the error region). The results presented in this paper were
obtained using the circular error region with $\delta y=0.05''$.

We minimize $\chi^2$ in the source plane to avoid having to search for
the image positions, thereby significantly reducing the computational
costs. Minimizing $\chi^2$ for the $\sim$10000 models takes $\sim5$
days CPU time on a SPARC 10 workstation.  We minimise $\chi^2$ using a
multi-dimensional Downhill Simplex method (Press et al 1992). Using
$\sigma_{\rm halo}$ and $f_{\rm halo}$ at $\chi^2_\rmn{min}$, we
calculate the time delay $\Delta t_\rmn{AB}$ between lensed images A
and B (Eqn. 16) and the magnifications $\mu_\rmn{A}$ and $\mu_\rmn{B}$
(Eqn. 14).

In total we have 5 fixed parameters (Table 3) and 5 constraints (flux
ratio, x$_1$ and x$_2$ positions of lensed images A and B, so
$\rmn{N}_\rmn{data}=5$).  We solve for the position of the source
($x_\rmn{s}$ and $y_\rmn{s}$), the velocity dispersion of the halo
$\sigma_\rmn{h}$ and the flattening of the halo $f_{\rm halo}$
($\rmn{N}_\rmn{pars}=4$). To avoid under constraining the system we
only solve for a total of 4 parameters ($\rmn{N}_\rmn{dof}=1$).

\subsection{Mass model parameters}

Below we will describe our choice for the fixed parameter space. For
each of the fixed parameters we take a broad range of values, in order
not to exclude possible models beforehand. All parameters are listed
in Table 3.

For the disk mass ${\rm M}_\rmn{disk}$ we take values of
(1.0-20.0)$\times10^{10}$ ${\rm M}_\odot$, spanning the range where
most `maximum disk' masses\footnote{The maximum disk mass is the
maximum mass one can attribute to the luminous mass (stellar+gas) of a
galaxy and still be in agreement with the observed HI rotation curve.}
of luminous disk galaxies lie (e.g. Broeils 1992). A `maximum disk'
mass maximizes the influence of the disk both on the dynamics and the
lensing properties. From the deconvolved NOT R-band image (Fig. 2) we
determine an axis ratio of the luminous disk of $\sim 0.3$. The HST
I-band image however shows that most emission (partly bulge) lies
clearly between the two QSO images, which are separated by
1.4$''$. This would imply an axis ratio $\la0.2$.  We therefore choose
to model the disk with axis ratios between 0.1 and 0.3, where an axis
ratio of 0.1 is typical for an edge-on disk (spiral) galaxy. The scale
length of the disk is hard to determine from either the NOT or the HST
images. Also the dust lane makes such a determination hard. We
therefore choose the large range of 1-16 kpc, knowing that the smaller
values are probably too small (as is also true for the small disk
masses). But these large parameter ranges make sure that we don't
underestimate the spread in the results that we obtain. We choose the
halo core radius between 0.05 and 3.2 kpc, depending on the choice of
mass model (NIE or MHP). The abcense of the central third image seems
to imply a high central surface density (small core radii; e.g.
Narayan \& Bartelman 1996), whereas rotation curve analyses seem to
imply close-to 'maximum disk' galaxies and therefore large halo core
radii (Rhee 1996; Broeils 1992).  As a central massive black hole in
lensing galaxies can significantly demagnify the central image, we
won't use the absence of this image as contraint on the core
radius. For the velocity dispersion of G2 we take the very large range
from 0 to 350 km/s. No velocity dispersion means that G2 has no
influence on the lensing properties. A redshift of G2 different from
G1 is approximately equivalent to decreasing its velocity
dispersion. In section 6.1.1 we will estimate a velocity dispersion
for G2 from its luminosity and use that to further constrain our
results.

\section{Results and analysis}
Here we describe the results of the minimization of $\chi^2$ for the
parameter space described in the previous section.  We assume that the
errors in the data have a Gaussian distribution.  In which case 95\%
of the $\chi^2$ distribution of each individual model has $\chi^2<4$
for N$_{\rm dof}=1$. In the analysis we therefore only consider models
with $\chi^2_\rmn{min}<4$ and $f_\rmn{halo}\le1$ (oblate halo models),
although in a sense the $\chi^2$ defined in the source plane is not
exactly equivalent to that in the source plane. A smaller cut-off
value (e.g. $\chi^2_\rmn{min}<1$) does not signifcantly effect the
results. Each model is weighted equally in the determination of the
average halo parameter values. Of the $\sim5000$ NIE halo models, 736
have $\chi^2_\rmn{min}<4$. Of the $\sim5000$ MHP halo models 182 reach
$\chi^2_\rmn{min}<4$.  This indicates that most combinations of the
fixed parameters cannot lead to a satisfactory fit to the
observational constraints. We will use all of the above models in the
determination of the halo flattening, halo velocity dispersion and
time delays between the two lensed images. We will also examine if
there are correlations between the fixed parameters (Table 3) and the
non-fixed parameters ($\sigma_{\rm halo}$ and $f_{\rm halo}$).

\subsection{Flattening of the halo}

\paragraph*{\underline{Analytical models}} First the flattening of the
total lensing mass distribution (luminous+\\dark) is calculated,
modeling G1 and its dark matter halo by a single Singular Isothermal
Ellipsoidal (SIE) mass distribution. We assume that
$\theta_\rmn{A}=1.14''$, $\theta_\rmn{B}=-0.25''$ and
$\theta_\rmn{G2}=-4.5''$, with the lensed images and galaxy G2 lying
on the x$_2$-axis of the coordinate system. Using the exact positions
of the lensed images and galaxy G2 gives essentially the same
results. We find the relation (Kormann et al. 1994):

\begin{equation} 
	\frac{|\theta_\rmn{A}-\theta_\rmn{B}|\rmn{D}_\rmn{d}}{\xi_0}=2\frac{\sqrt{f}}
	{\sqrt{1-f^2}}\mbox{arcsinh} \left(\frac{\sqrt{1-f^2}}{f}\right), 
\end{equation}
with
\begin{equation} 
	\xi_0=4\pi \left(\frac{\sigma}{c}\right)^2\frac{\rmn{D}_\rmn{d} \rmn{D}_\rmn{ds}}
	{\rmn{D}_\rmn{s}}, 
\end{equation}
independent of the presence of G2.  In Eqns. 25 and 26, $f$ is the
axis ratio $(b/a)_\Sigma = (c/a)_\rho$ of the SIE mass distribution
($f$ equals the oblateness for an edge-on mass distribution), $\sigma$
is the velocity dispersion of the combined luminous+dark mass
distribution, $\xi_0$ is the Einstein radius, $\rmn{D}_\rmn{s}$,
$\rmn{D}_\rmn{d}$ and $\rmn{D}_\rmn{ds}$ are the angular diameter
distances between respectively observer-source, observer-lens and
lens-source. The magnification ratio is given by

\begin{equation} 
	\frac{\mu_\rmn{A}}{\mu_\rmn{B}}=\frac{\left({1-\frac{\xi_0}{\rmn{D}_\rmn{d}}
	[\frac{\sqrt{f}}
	   {|\theta_\rmn{B}|}}-
           \left(\frac{\sigma_\rmn{G2}}{\sigma}\right)^2\cdot
	   \frac{1}{|\theta_\rmn{B}-\theta_\rmn{G2}|}] \right)}
	   {\left({1-\frac{\xi_0}{\rmn{D}_\rmn{d}}[\frac{\sqrt{f}}{|\theta_\rmn{A}|}}-
           \left(\frac{\sigma_\rmn{G2}}{\sigma}\right)^2\cdot
	   \frac{1}{|\theta_\rmn{A}-\theta_\rmn{G2}|}] \right)},
\end{equation}
which can be reduced to
\begin{equation} 
	\frac{\mu_\rmn{A}}{\mu_\rmn{B}}=
		\left[\frac{1-2(\kappa^{\rm B}+\kappa_\rmn{G2}^{\rm B})}
                          {1-2(\kappa^{\rm A}+\kappa_\rmn{G2}^{\rm A})}\right],
\end{equation}
where $\kappa^\rmn{A/B}$ and $\kappa_\rmn{G2}^\rmn{A/B}$ are the
dimensionless surface densities (Eqn. 2) at the positions of the
lensed images A and B. In Fig. 5 we plot the magnification ratio (flux
ratio) given by Eqn. 28 as function of $f$ and $\sigma$.  Using
${\mu_\rmn{A}}/{\mu_\rmn{B}}=-1.25$, where the minus sign comes from
the parity difference between the lensed images, we can solve $f$ and
$\sigma$ from Eqns. 25-28.  If a non-zero core radius is introduced in
the mass distribution we find that for a constant flux ratio
($|{\mu_\rmn{A}}/{\mu_\rmn{B}}|$) the corresponding flattening $f$ of
the mass distribution increases. This implies that Eqn. 28 gives a
lower limit on the flattening. Also the exclusion in Eqn. 28 of the
flat luminous mass distribution implies that $f$ is smaller than the
flattening $f_\rmn{halo}$ of the dark matter halo, because the
combined luminous+dark matter is flatter then the dark matter (if
$f_\rmn{disk}<f_\rmn{halo}$).

\paragraph*{\underline{Numerical models}} One can compare the
analytical results for the mass flattening with what is found from the
more detailed numerical models where the disk, bulge and core radius
of the halo are taken in to account. We examined the correlations
between $f_\rmn{halo}$ and the fixed parameters in Table 3 and only
found a clear anticorrelation with $\sigma_\rmn{G2}$.  In Fig. 6 we
plot the flattening $f_\rmn{halo}$ of the NIE halo models against
$\sigma_\rmn{G2}$ for all models with $\chi^2_\rmn{min}<4$.  If
$\sigma_\rmn{G2}$ increases $f_\rmn{halo}$ decreases, which is
qualitatively in agreement with the analysis given above (Fig. 5). The
solutions of $f_\rmn{halo}$ from these more detailed models are
systematically slightly higher than the values found for $f$ from
Eqn. 28, especially for the smaller values of $\sigma_\rmn{G2}$.

It is evident from Fig. 6 that $\sigma_\rmn{G2}$ is a very important
parameter in the determination of limits on $f_\rmn{halo}$. In the
next section constraints on $\sigma_\rmn{G2}$ will be derived, which
are subsequently used to constrain $f_\rmn{halo}$, using both
analytical and more detailed numerical models.

\subsubsection{Velocity dispersion of G2 and limits on $f_\rmn{halo}$}

To estimate the velocity dispersion $\sigma_\rmn{G2}$ we use the
Tully-Fisher relation and the V-magnitude of galaxy G2. From our HST
V-band observation we find a V-magnitude of 20.6 for G2, consistent
with the results of JH97. From Coleman, Wu \& Weedman (1980) we find a
K correction of 0.6 and 1.2 for (B-V) (if G2 is a Sbc galaxy at
$z_\rmn{d}=0.415$; From JH97 (B-V)$\sim1.4$). The luminosity is given
by
 
\begin{equation}
	{\rm L}_\rmn{B}=10^{0.4({\rm M}_{{B}\odot}-{V}-
		({B}-{V})+\rmn{DM}+{K})}~\rmn{L}_{\odot},
\end{equation}
where $\rmn{M}_{\rmn{B}\odot}=5.48$ is the total solar B magnitude and
DM is the distance modulus. Using $z_\rmn{d}=0.415$ we find
DM=42.2-$5\cdot\log(h_{50})$.  The luminosity of G2 in B is then $
{L}^\rmn{G2}_\rmn{B}\sim4\cdot10^{10}/h_{50}^2~~~\rmn{L}_{\odot}.$
Using this luminosity we find from Rhee (1996) that
$\log{v_f}\approx2.3$ ($h_{50}=1$), where $v_f$ is the rotation
velocity of the HI gas in the flat part of the rotation curve. Using
the relation $v_f\approx \sqrt{2}\sigma$ we find that the velocity
dispersion of G2 must be $\sigma_\rmn{G2}\sim140$ km/s, under the
assumption that G2 can be described by a Singular Isothermal Sphere
and that the local ($z=0$) Tully-Fisher relation holds at a redshift
of z=0.4.  To find a strong upper limit on the velocity dispersion of
G2 we go through the sample of luminous spiral galaxies of Broeils
(1992) and Rhee (1996). No galaxies with $\log{v_{\rmn{max}}}>2.54$
are found.  This implies an upper limit on $v_{\rmn{max}}$ of 350
km/s, or an upper limit of about 250 km/s on the velocity disperion of
G2 ($\sigma\sim v_{\rm f}/\sqrt{2}$). We find from Figure 6 that
$\sigma_\rmn{G2}\sim140$ km/s would imply an almost spherical halo
with $f_{\rm halo}\ga0.8$. The more stringent upper limit of
$\sigma_\rmn{G2}\la 250$ km/s gives a lower limit of $f_{\rm
halo}\ga0.5$.  This compares well with the lower limit
$f_\rmn{halo}\ga0.4$ for the same range of velocity dispersions of G2,
which we find from equation 28 and Figure 5.

Moreover, Fig. 6 shows $f_\rmn{halo}$ plotted against
$\sigma_\rmn{G2}$ for all MHP halo models with
$\chi^2_\rmn{min}<4$. For the range $\sigma_\rmn{G2}=140-250$ km/s we
find $f_\rmn{halo}\ga0.50$, identical to the limit on $f_\rmn{halo}$
from the NIE halo models.

Except for the strong anticorrelation between $\sigma_\rmn{G2}$ and
$f_\rmn{halo}$, no other significant correlations are found. The
spread in $f_\rmn{halo}$ for fixed values of $\sigma_\rmn{G2}$ appears
therefore to result mainly from the spread in image positions and flux
ratio.

So for both halo models (NIE and MHP) we find a lower limit
$f_\rmn{halo}\ga0.5$ on the halo flattening, using the range of
velocity dispersion of G2, $\sigma_\rmn{G2}=140-250$ km/s.  We also
see that the lower limit on $f_\rmn{halo}$ is not strongly dependent
on the chosen halo model. We recalculated a subsample of the models,
using different starting values of $f_\rmn{halo}$ and
$\sigma_\rmn{halo}$, finding the same solutions for both parameters
for $\chi^2_{\rm min}$. The agreement between the lower limits on
$f_\rmn{halo}$ between the NIE and MHP halos are therefore not an
artifact of the initial values of both parameters.  Moreover one would
expect to find lower values for $f_{\rm halo}$ for the more centrally
concentrated MHP models, compared with the NIE halo models. We find
that the low $f_\rmn{halo}$ solutions for the MHP halo are mostly
models with $r_{\rm c}\ge1.6$ kpc. These models are not very centrally
concentrated and therefore allow for larger values of $f_{\rm halo}$.
The lower limits therefore appear genuine and not artificial.

From the analytical models (Figure 5) we find that the flux ratio
$r_\rmn{AB}$ strongly constraints the flattening of the SIE mass
distribution.  An error of $\pm0.05''$ in the distance of images A and
B to the lens center of G1 gives an error of 0.1 in the
flattening. Using the same range of $\sigma_\rmn{G2}=140-250$ km/s, we
find a lower limit on the combined mass distribution (halo+disk+bulge)
of $0.40\pm0.1$.  Because the disk is much flatter than the halo and
on first sight much more massive than the bulge, the same lower limit
applies to the halo. Moreover a non-zero core radius will increase the
value of $f_{\rm halo}$.

Moreover, we have tried modeling B1600+434, using only the disk and
bulge components.  If we constrain the bulge mass, we find that the
axis ratio of the disk typically will increase to $f_{\rm
disk}\ga0.5$, larger than the observed limit of 0.3. On the other
hand, if the disk flattening is constraint to $\le 0.3$, we find
solutions that required extremely large bulge masses (${\rm M}_{\rm
bulge}> {\rm M}_{\rm disk}$).  This again supports the need for a mass
component rounder than the disk, but much more massive than the bulge.

\subsubsection{Critical curves and caustics}

In Fig. 7 the critical and caustic curves for two distinct NIE halo
mass models are shown. Both models give good fits to the observed
image positions and flux ratio, but the critical and caustic structure
is quite different.

Fig. 7a and 7b both show a mass model with an almost spherical halo
and velocity dispersion near 200 km/s. The difference between both
models is the mass and flattening (axis ratio) of the disk. Also the
velocity dispersion of G2 is different between both models. It is
already evident from these two models that the presence of a flat
stellar mass distribution can significantly alter the critical and
caustic structure of the lens and still be in agreement with the
observed image postions and flux ratio. The model shown in Fig. 7a has
a larger 5-image cross section and one expects therefore a different
ratio of 5 to 3-image systems between these models. A precise model of
the stellar mass distribution is therefore needed to understand
exactly the statistical properties of gravitational lenses with highly
flattened luminous mass distributions. This is particulary evident in
the case of B1600+434.  A more detailed analysis of the statistics of
spiral galaxy lenses can be found in Koopmans \& Nair (1997).

All our models place the source position close to the radial caustic,
whereas Maller et al. (1997) find a source position close to the
tangential caustic. This is the result of a difference in adopted lens
center. The difference between our and their lens center is $\sim
0.2''$, whereas our HST observations allow only a $\sim0.05''$
difference. If we move our lens center to the position found by Maller
et al. (1997), we find that the source position moves from the radial
to the tangential caustic.  Our observations allow a very small shift
in the lens center, but much smaller than $0.2''$.  This shift would
move the source postion slighty away from the radial caustic.

\subsubsection{The importance of the redshift of G2}

From Kochanek \& Apostolakis (1988) we find that two lenses interact
significantly if the transverse separation between the two lenses is
$\la4$ times the radius of the outer critical curve. From Fig. 7 we
find that the radius of the outer critical curve in the direction of
G2 is $\sim0.7''$ (approximately the Einstein radius or half the image
separation).  We see that $4\times0.7''=2.8''<4.5''$, where the
separation between G1 and G2 is $\sim4.5''$. Although the critical
curves of G1 will be somewhat distorted (Figure 7), galaxy G2 can in
first order be approximated by a perturbing external shear. The
strength of this shear is a function of both the redshift and velocity
dispersion of G2. According to Kochanek \& Apostolakis (1988) both
lenses will work together most efficiently, if they are at the same
intermediate redshift. Changing the redshift of G2, will therefore
decrease the strength of this perturbing shear, which in first order
is equivalent to decreasing the velocity dispersion of G2. This effect
only becomes significant when $z_{\rm G2}$ is outside the range of
$z_{\rm G1} \pm 0.3 = 0.1-0.7$.  Furthermore a much higher redshift
for G2 can change the geometrical part of the timedelay surface,
changing the expected time delay. However, the colors of G2 seem to
indicate that a redshift smaller than 0.4 is more likely than a
redshift larger than 0.4. The B-V and V-I colors from JH97 and our
F555W-F814W colors indicate that a redshift of 0.8 for G2 cannot be
accomodated by either E/S0 or spiral galaxies. A mucher larger
redshift than 0.4 is therefore unlikely for G2.  The effect of G2 on
the timedelay will therefore in first order be included in our range
of velocity dispersion for G2 and as we will see the effect of G2 will
be marginal.

\subsection{Velocity dispersion of the halo}

In this section we will describe our results for the velocity
dispersion of the NIE halo. We only give the results of
$\sigma_\rmn{halo}$ for the NIE halo models, because it can directly
be related to rotation curve observations of spiral galaxies.

\paragraph*{Velocity dispersions and mass}

From Eqns. 25 and 26 we find that $\sigma_\rmn{halo}\approx200$ km/s,
nearly independent of the flattening of the halo
($(\theta_\rmn{A}-\theta_\rmn{B})\rmn{D}_\rmn{d}/\xi_0\approx2$ for
$f\ga0.1$). This velocity dispersion implies a total mass inside the
Einstein radius ${\rm M}(\xi<\xi_0)\approx 1.3\cdot10^{11}$M${}_\odot$
($h_{50}=1$).  In Fig. 8 we plot the histogram of $\sigma_\rmn{halo}$
for the NIE halo models with $\chi^2_\rmn{min}<4$. It appears that
$\sigma_\rmn{halo}$ is restricted to a small range of values. Simply
taking the average for all models with $\chi^2_\rmn{min}<4$ we find
$\sigma_\rmn{halo}=190\pm15$ km/s, consistent with what we found
analytically. This velocity dispersion gives a rotation velocity
outside the optical disk of $v_\rmn{f}\approx 270$ km/s, comparable to
rotation velocities found for large luminous spiral galaxies
(e.g. Broeils 1992). Restricting to the larger and more massive models
with $h_{\rm disk}\ge 8$ kpc and ${\rm M}_\rmn{disk}\ge
5\cdot10^{10}\;{\rm M}_\odot$, we find only a slight decrease in the
halo velocity dispersion, $\sigma_\rmn{halo}=180\pm15$ km/s. Both
distributions have a wing towards to lower velocity dispersions
(Figure 8).

Using the WFPC2 F814W magnitude of 20.2 and the V-F814W$\approx$2.0
magnitude for a Sab galaxy at $z\sim0.5$ (Fukugita et al. 1995), we
expect a lower limit (due to dust obscuration) on the V magnitude of
22.2, close to the 22.0 found by JH97. This results in a luminosity of
${\rm L}_\rmn{B}^{\rm G1} \sim10^{10}~\rmn{L}_{\odot}$. From Rhee
(1996) we expect that an Sab galaxy at this redshift should be
$\sim2.5$ magnitudes brighter in B. This difference again indicates
the presence of a large amount of obscuring dust. Calculating a
sensible mass-to-light ratio is therefore difficult in the inner parts
of G1. Moreover the presence of image B in the bulge of G1 makes this
even harder. Because of the expected few magnitudes of extinction, the
M/L ratio of 51 in B (JH97) could be $\sim5-10$ times smaller. The
ratio would then be in the range of spiral galaxies.

\paragraph*{Correlations of $\sigma_\rmn{halo}$ with disk mass}

We find an anti-correlations between the $\sigma_\rmn{halo}$ and ${\rm
M}_\rmn{disk}$ (Fig. 9). The separation between the two lensed images
is a strong function of the mass inside the Einstein radius.  This
means that an increase in ${\rm M}_{\rmn{disk}}$ increases the mass
inside the Einstein radius (between the lensed images). This increase
in mass must be compensated by the other mass component, the dark
halo, hence $\sigma_\rmn{halo}$ decreases. For the massive disk models
(${\rm M}_{\rmn{disk}}\ga10^{11}$M${}_\odot$) we find that only models
with large exponential scale lengths\footnote{e.g Most of the disk
mass is outside the Einstein radius.} ($\ge4$ kpc), give solutions
with $\chi^2_\rmn{min}<4$ and that $\sigma_\rmn{halo}$ does not drop
below $\sim150$ km/s (for $\sigma_{\rm G2}\la 250$ km/s). We conclude
that even very high disk masses (masses larger than that needed for
the image splitting) do not significantly reduce the velocity
dispersion of the halo. Indeed a massive halo is in needed to fit the
observations. Also a small anti-correlation is found between
$\sigma_\rmn{halo}$ and $\sigma_\rmn{G2}$, which could be explained by
the model trying the match the flux ratio when changing
$\sigma_\rmn{G2}$.

\subsection{Time delay}

For the SIE mass distribution and under the same assumptions as in
section 6.1.1 we find for the time delay between the lensed images
(Kormann et al. 1994):

\begin{equation} 
	\Delta t_\rmn{AB}=\frac{\xi_0}{c}\frac{\rmn{D}_\rmn{s}}{\rmn{D}_\rmn{ds}} (1+z_\rmn{d}) 
	\frac{\sqrt{f}}{\sqrt{1-f^2}}
        \cdot \mbox{arccosh}(\frac{1}{f}) (|\theta_\rmn{A}|-|\theta_\rmn{B}|). 
\end{equation}
Eqn. 30 is equivalent to the time delay without the presence of G2 in
the mass model.  This indicates that one does not expect a very large
influence of G2 on the time delay between the lensed images. Using
Eqn. 25 we can reduce Eqn. 30 to

\begin{equation} 
	\Delta t_\rmn{AB}=\frac{1}{2c}\left(\frac{\rmn{D}_\rmn{d} \rmn{D}_\rmn{s}}
	{\rmn{D}_\rmn{ds}}\right)
	(1+z_\rmn{d})|\theta_\rmn{A}-\theta_\rmn{B}|
	         (|\theta_\rmn{A}|-|\theta_\rmn{B}|)
\end{equation}
For $z_\rmn{d}=0.415$, $z_\rmn{s}=1.59$, $\theta_\rmn{A}=1.14''$,
$\theta_\rmn{B}=-0.25''$ and an error in $\theta_\rmn{AB}\sim0.05''$
we find a time delay $\Delta t_\rmn{AB}=(57\pm7)/h_{50}$ days. We will
now compare this predicted time delay with the time delays found from
our numerical models.

In Fig. 10 we plot the histogram of the time delay for all numerical
models with $\chi^2_\rmn{min}<4$. We see in this figure that the time
delay depends only weakly on variations in the input model
parameters. Taking the average of all time delays with
$\chi^2_\rmn{min}<4$ we find $\Delta
t_\rmn{AB}^{\rmn{NIE}}=(54\pm3)/h_{50}$ days.  Using only the larger
and more massive models with $h_{\rm disk}\ge 8$ kpc and ${\rm
M}_\rmn{disk}\ge 5\cdot10^{10}{\rm M}_\odot$, we find $\Delta
t_\rmn{AB}^{\rmn{NIE}}=(53\pm3)/h_{50}$ days. This is in excellent
agreement with the time delay we found from our simple analysis,
indicating the stability of the time delay against changes in the
fixed NIE model parameters.

In Fig. 10 we also plotted the histogram of the time delay for all MHP
models with $\chi^2_\rmn{min}<4$.  Taking the average of these time
delays, we find $\Delta t_\rmn{AB}^{\rmn{MHP}}=(70\pm4)/h_{50}$ days.
For the larger more massive models we find $\Delta
t_\rmn{AB}^{\rmn{MHP}}=(68\pm2)/h_{50}$ days.  Also for the MHP halo
models a small spread in the time delay is found.  It is therefore
clear, as in the case of the NIE halo model, that changes in the fixed
model parameters do not severely influence the time delay between the
lensed images. We have not calculated the time delay for the MHP mass
models analytically .

The difference in time delays between the MHP and NIE halo models are
$\sim30\%$, because the more centrally concentrated MHP mass
distribution gives rise to a much larger potential timedelay between
images A and B. Although the flat rotation curves of spiral galaxies
seem to favor a NIE mass distribution, we do not know the distribution
of mass in the z-direction very well. More contraints on the mass
distribution are therefore needed (e.g VLBI structure in the images A
and B).

\subsection{Magnification of images A and B}

We have calculated the magnification of the lensed images for every
model with $\chi^2_\rmn{min}<4$. The averages of the magnifications
and flux ratios for the NIE and MHP halo models are listed in Table
3. We see that the average magnifications for both halo models are not
very large, but typical for a two image system. One also notes that
the magnification from the MHP halo models is smaller than the
magnification from the NIE halo models. The difference between the
magnifications is about 33\%, which results in a different calculated
absolute magnitude for the lensed source for the two (NIE and MHP)
halo models.

\section{Conclusions}

New HST and NOT observations of the GL system B1600+434 strongly
suggest that the lensing galaxy in this system is an edge-on spiral
galaxy. Because the system is nearly edge-on, we can use the lensing
properties of this system to constrain the velocity dispersion and
oblateness (flattening) of the dark matter halo around the lensing
spiral galaxy. This sytem is unique in the sense that for the first
time gravitational lensing has been used to constrain the dark matter
distribution around an individual spiral galaxy. Moreover, the lensed
QSO is highly variable both in the radio and the optical and can
therefore be used to determine the time delay between the two lensed
images. This time delay gives us either H$_0$, once the mass model has
been well constrained, or an extra constraint on the mass model, once
H$_0$ has been contrained from other GL systems (e.g. B0218+357,
B1608+656, etc).

\subsection{Flattening of the halo}

From detailed numerical modeling we find a lower limit of
$f_\rmn{halo}\ga0.50$ on the axis ratio of both the NIE and MHP halo
mass distribution around the edge-on spiral galaxy lens in B1600+434.

When we use a simplified analytical SIE mass model we do not find a
surface mass distribution significantly flatter than $\sim0.4$.  Using
a core radius $r_c>0$ increases this limit. This lower limit is
significantly larger than the typical flattening ($\sim0.1$) of the
luminous stellar component of spiral galaxies and also larger than the
upper limit on the flattening of G1 ($\la 0.3$) from the HST and NOT
observations.  We conclude that the halo dark matter around G1 is not
as flat as the luminous stellar component (or gas). This implies that
the suggestion by Pfenniger at al. (1994), that dark matter could be
cold molecular hydrogen associated with HI gas, is inconsistent with
our results for G1. In that case one would expect to find a halo
flattening in the same order or smaller than the luminous stellar
component ($f_{\rmn{halo}}\le f_\rmn{disk}$), which we clearly do not
find.

\subsection{Velocity dispersion and mass of the halo}

The average velocity dispersion over all models of the NIE halo is
found to be $190\pm15$ km/s, which gives a rotation velocity outside
the optical disk of $\sim$270 km/s ($f_\rmn{halo}\sim1$), consistent
with luminous spiral galaxies (e.g. Broeils 1992). For large disk
masses ($10-20\cdot 10^{10}$ M$_\odot$) the velocity dispersion of the
halo decreases, but never drops below $\sim 150$ km/s, even for disk
masses larger than the mass needed for the image splitting. This
indicates that a massive halo is needed around the much flatter
luminous stellar component to fit the observed image positions and
flux ratio.

The mass inside the Einstein radius is $\sim1.3\cdot
10^{11}$M${}_\odot$ of which at least half (if $\sigma_\rmn{halo}\ga
150$ km/s) can be attributed to the halo.  Using the Tully-Fisher
relation and the observed $10^{10}~\rmn{L}_{\odot}$ in the B band, we
suspect some 2.5 magnitudes of extinction in the B band. If this is
the case, the mass-to-light ratio of 51 in B (JH97) would be reduced
to $\sim5$, consistent with spiral galaxy mass-to-light ratios.

\subsection{Time delay}

The time delays found for the different halo models are quite
different and do not significantly depend on the presence of G2.  The
NIE halo model gives in a time delay of $\Delta
t_\rmn{A/B}^{\rmn{NIE}}=(54\pm3)/h_{50}~~\rmn{days}$, whereas the MHP
halo model gives $\Delta
t_\rmn{A/B}^{\rmn{MPH}}=(70\pm4)/h_{50}~~\rmn{days}$.  These delays
decrease by only a few percent if only models with $h_{\rm disk}\ge 8$
kpc and ${\rm M}_\rmn{disk}\ge 5\cdot10^{10}\;{\rm M}_\odot$ are
used. It is clear that more observations are necessary to constrain
the mass model and discriminate between different halo mass models.
Because flat rotation curves of spiral galaxies seem to point at a NIE
halo, the time delay from the NIE halo model is probably closest to
the actual time delay. The time delay from the MHP can still be used
to put an upper limit on H${}_0$.

\section*{Acknowledgments}

L.V.E. Koopmans and A.G. de Bruyn acknowledge the support from an NWO
program subsidy (grant number 781-76-101). We would like to thank
Sunita Nair for the many valuable discussions about the modeling of
lens systems and many more aspects of lensing. We furthermore thank
Chris Fassnacht for kindly supplying us with the redshifts of the
lensing galaxy and the lensed source and Penny Sackett for giving a
number of constructive comments and suggestions. This research used
observations with the Hubble Space Telescope, obtained at the Space
Telescope Science Institute, which is operated by Associated
Universities for Research in Astronomy Inc. under NASA contract
NAS5-26555. MERLIN is a national facility operated by the University
of Manchester on behalf of SERC. The Westerbork Synthesis Radio
Telescope (WSRT) is operated by the Netherlands Foundation for
Research in Astronomy (ASTRON) with the financial support from the
Netherlands Organization for Scientific Research (NWO).


\newpage
\pagestyle{empty}

\noindent
{\bf \huge Captions:}\\

{
\medskip
\noindent
{\bf Fig.1 :} HST I band image. North is up, east is left.\\

\medskip
\noindent
{\bf Fig.2 :} {\bf Upper} : Nordic Optical Telescope (NOT) B band image, {\bf middle} : 
NOT R band-image, {\bf lower} : deconvolved NOT R band image. 1$''$ corresponds to
a physical size of 6.5 kpc at redshift $z_l=0.415$ and $h_{50}=1$.\\

\medskip
\noindent
{\bf Fig.3 :} MERLIN 5 GHZ observation of B1600+434 (March 14, 1995).\\

\medskip
\noindent
{\bf Fig.4 :} Total flux density of the two QSO images at 21 cm radio continuum measured with
the WSRT as a function of time from April 8, 1996. The error bars indicate the flux 
error in fitting the model to the visibilities.\\

\medskip
\noindent
{\bf Fig.5 :} The flux ratio $|{\mu_\rmn{A}}/{\mu_\rmn{B}}|$ plotted as function of $f$ for 
$\theta_\rmn{B}=-0.20''\mbox{(dot)}, -0.25''\mbox{(dash)},$ $-0.30''\mbox{(long dash)}$ 
and for $(\sigma_\rmn{G2}/\sigma)=0.5, 1.0, 1.5, 2.0$, using a simple SIE surface 
mass model to describe G1 and SIS mass model
for galaxy G2. The horizontal solid line gives the flux ratio of 1.25, the 
dotted lines border the region 1.15-1.35.\\ 

\medskip
\noindent
{\bf Fig.6 :} The flattening parameter $f_\rmn{halo}$ plotted against $\sigma_\rmn{G2}$ for
the NIE (square) and MHP halo models (triangle) with $\chi^2_\rmn{min}<4$. The velocity
dispersion $\sigma_\rmn{G2}$ has been shifted by -3 km/s and +3 km/s for
respectively the NIE and MHP halo models.\\

\medskip
\noindent
{\bf Fig.7 :} The critical lines (dashed) and caustics (solid) for two distinct NIE halo mass models
of B1600+434 with an NIE halo. The small plusses give the model positions of the lensed images. 
The circles give the positions of the lensed images listed in Table 1. The larger cross indicates
the calculated position of the source in the source plane. The large shift 
in the position of the caustics is  the result of the presence of G2. 
In the upper figure (a) we see a model with $r_\rmn{c}=0.40$ kpc, 
$f_\rmn{halo}=0.89$, $\sigma_\rmn{halo}=199$ km/s, $h_\rmn{disk}=8$ kpc, $f_\rmn{disk}=0.1$, 
$\rmn{M}_\rmn{disk}=5.0\times10^{10} \rmn{M}_{\odot}$ and $\sigma_\rmn{G2}=150$ km/s.
In the lower figure (b) we see a model with $r_\rmn{c}=0.40$ kpc, 
$f_\rmn{halo}=0.95$, $\sigma_\rmn{halo}=193$ km/s, $h_\rmn{disk}=8$ kpc, $f_\rmn{disk}=0.3$, 
$\rmn{M}_\rmn{disk}=1.0\times10^{11} \rmn{M}_{\odot}$ and $\sigma_\rmn{G2}=200$ km/s.\\

\medskip
\noindent
{\bf Fig.8 :} Histogram (solid) of the velocity dispersion of the NIE halo for 
$\chi^2_\rmn{min}<4$. The dashed histogram are those models with
${\rm M}_\rmn{disk}\ge 5\cdot10^{10}\;{\rm M}_\odot$ and $h_{\rm disk}\ge 8$ kpc.\\

\medskip
\noindent
{\bf Fig.9 :} Correlation between $\sigma_\rmn{halo}$ and 
M${}_{\rmn{disk}}$ for the NIE halo model.\\

\medskip
\noindent
{\bf Fig.10 :} Histogram of the NIE (upper) and MHP (lower) model time delays for 
$\chi^2_\rmn{min}<4$. The dashed histograms are those models with
${M}_\rmn{disk}\ge 5\cdot10^{10}{\rm M}_\odot$ and $h_{\rm disk}\ge 8$ kpc.\\


\newpage
\pagestyle{empty}

\begin{figure}
\begin{center}
\leavevmode
\vspace{3cm}
\vbox{%
\epsfxsize=15cm
\epsffile{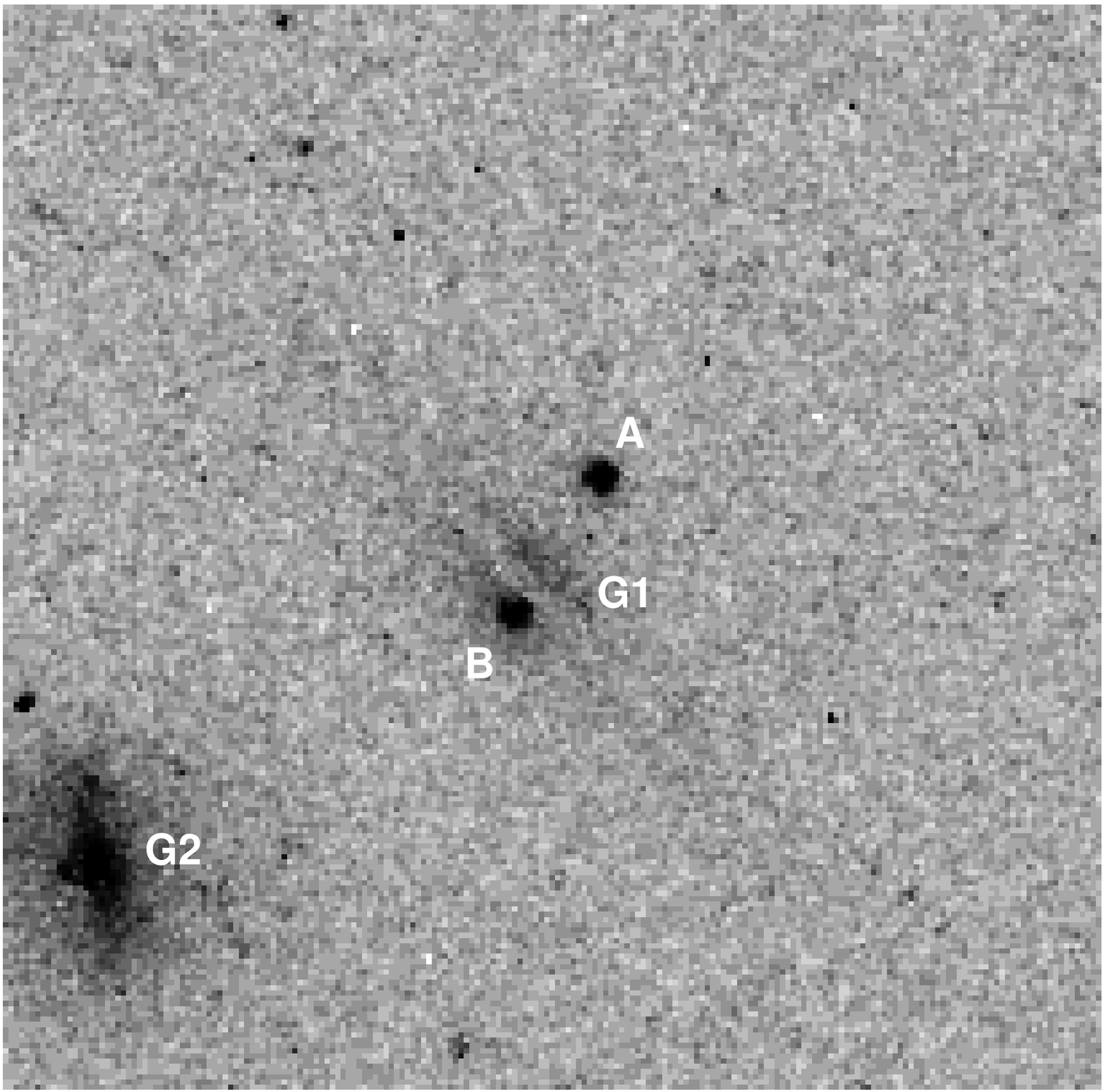}}
\vspace{3cm}
\end{center}
\end{figure}

\begin{figure}
\begin{center}
\leavevmode
\vbox{%
\epsfxsize=9.0cm
\epsffile{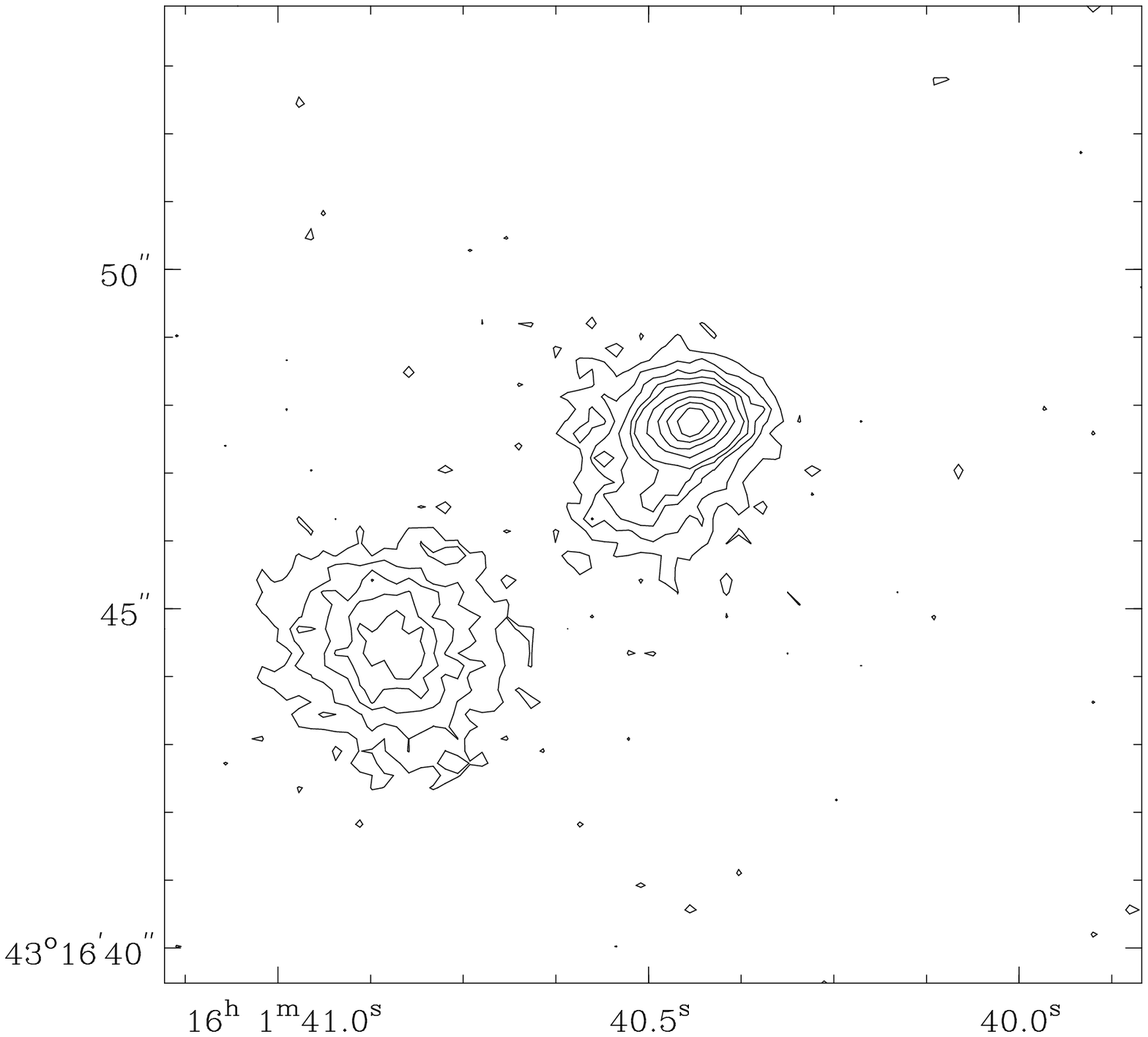}
\epsfxsize=9.0cm
\epsffile{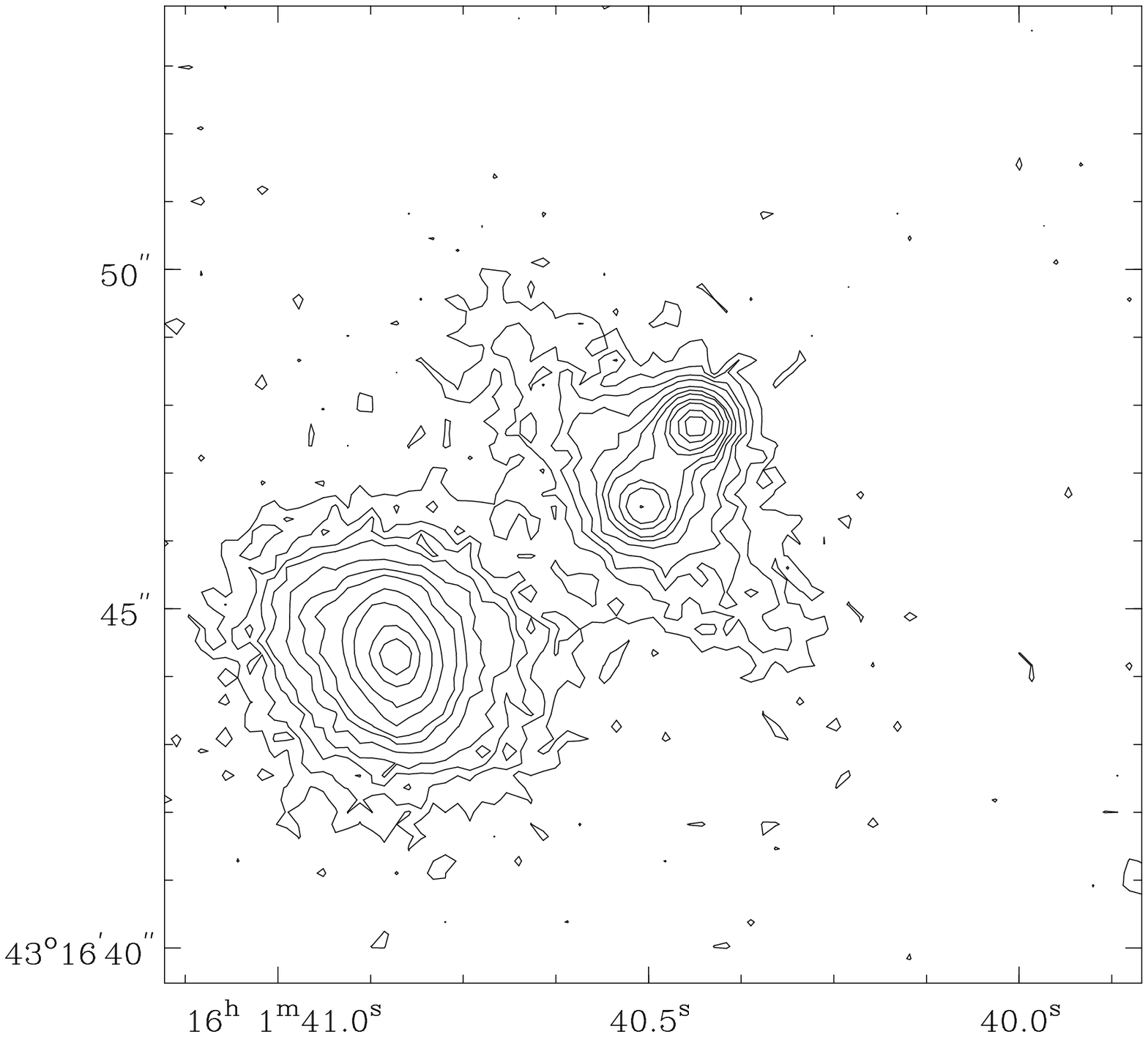}
\epsfxsize=9.0cm
\epsffile{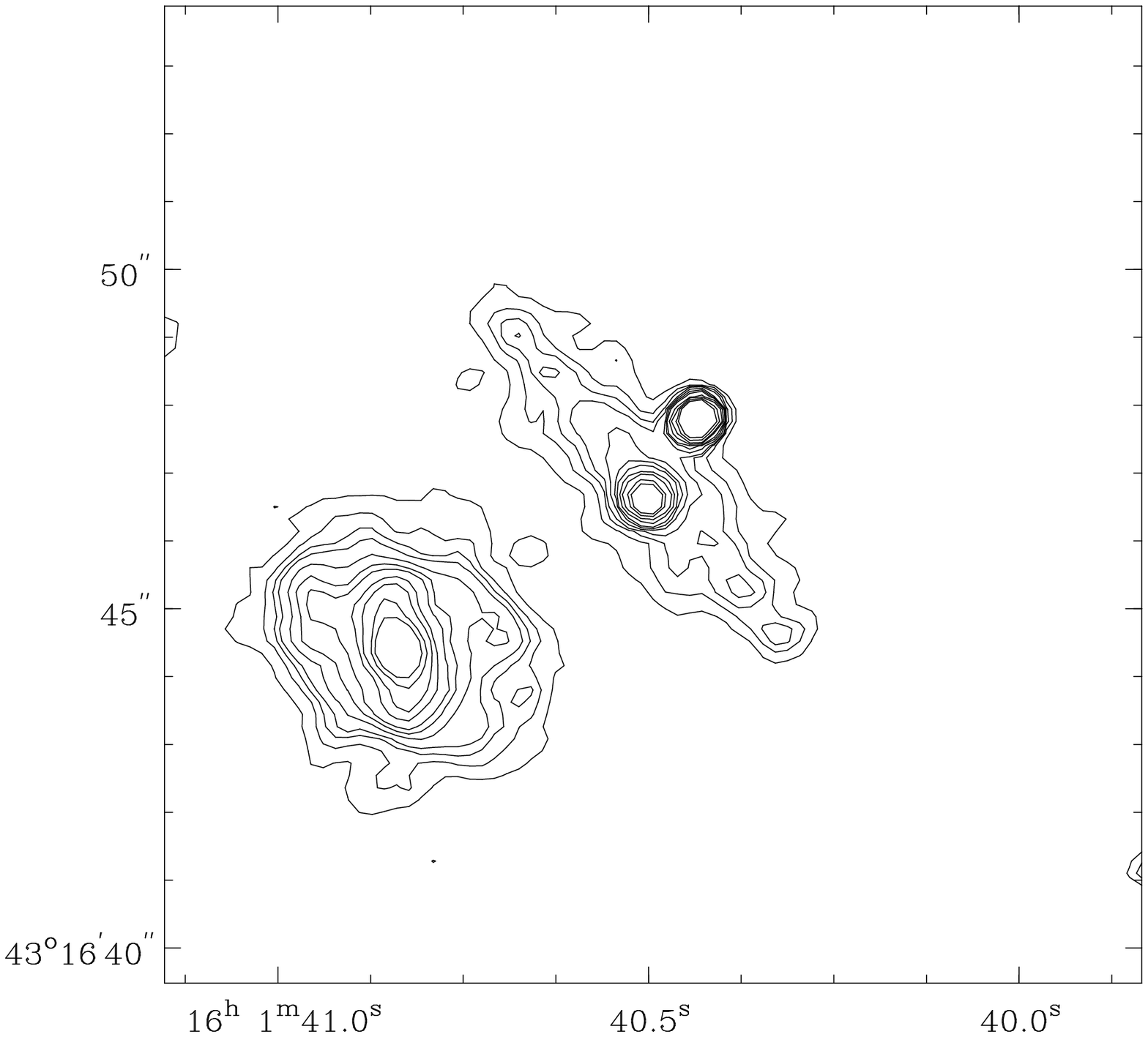}}
\end{center}
\end{figure}

\begin{figure}
\begin{center}
\leavevmode
\vbox{%
\epsfxsize=15.0cm
\epsffile{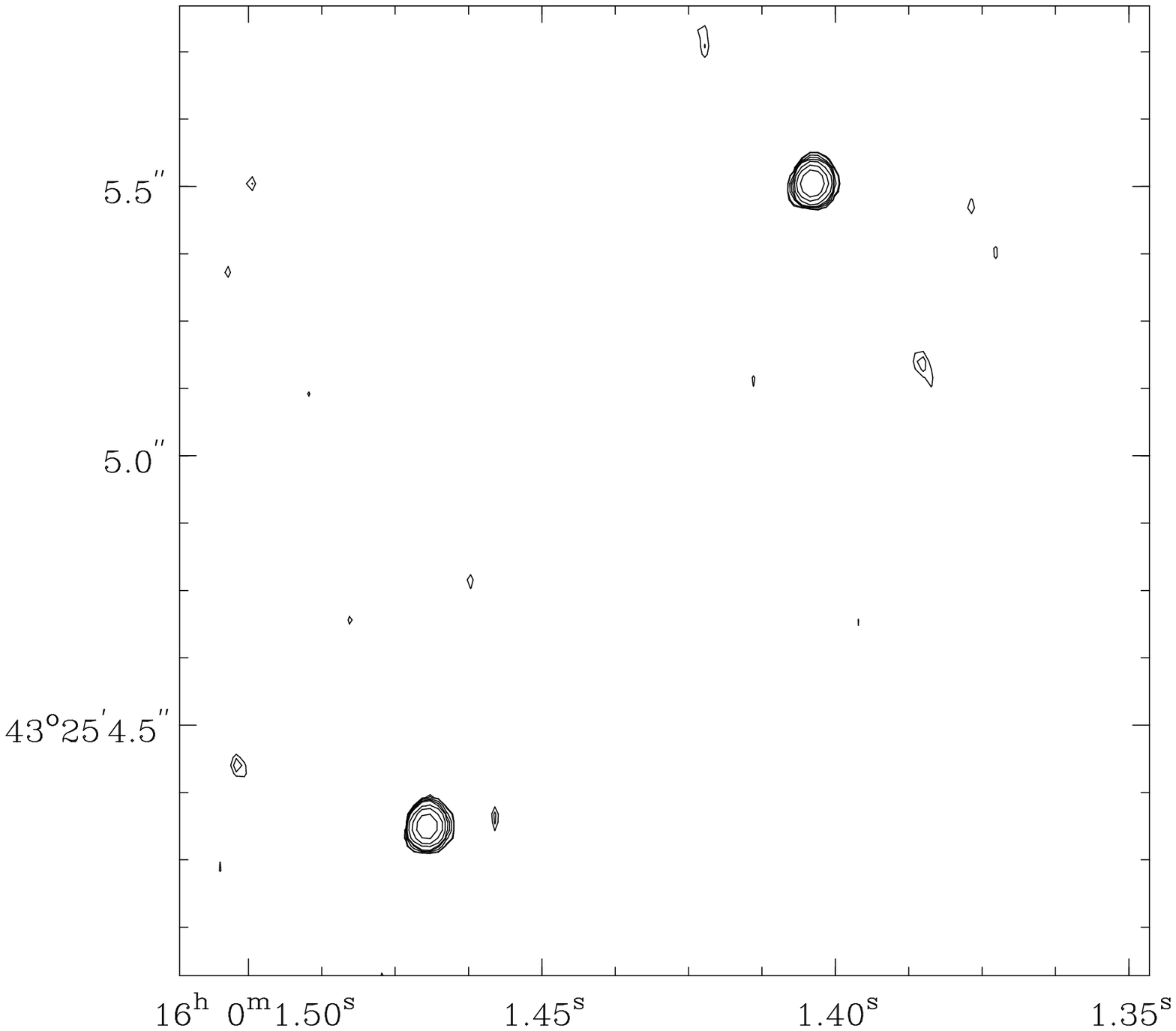}}
\end{center}
\end{figure}

\begin{figure}
\begin{center}
\leavevmode
\hspace{0.5cm}
\vbox{%
\epsfxsize=15cm
\epsffile{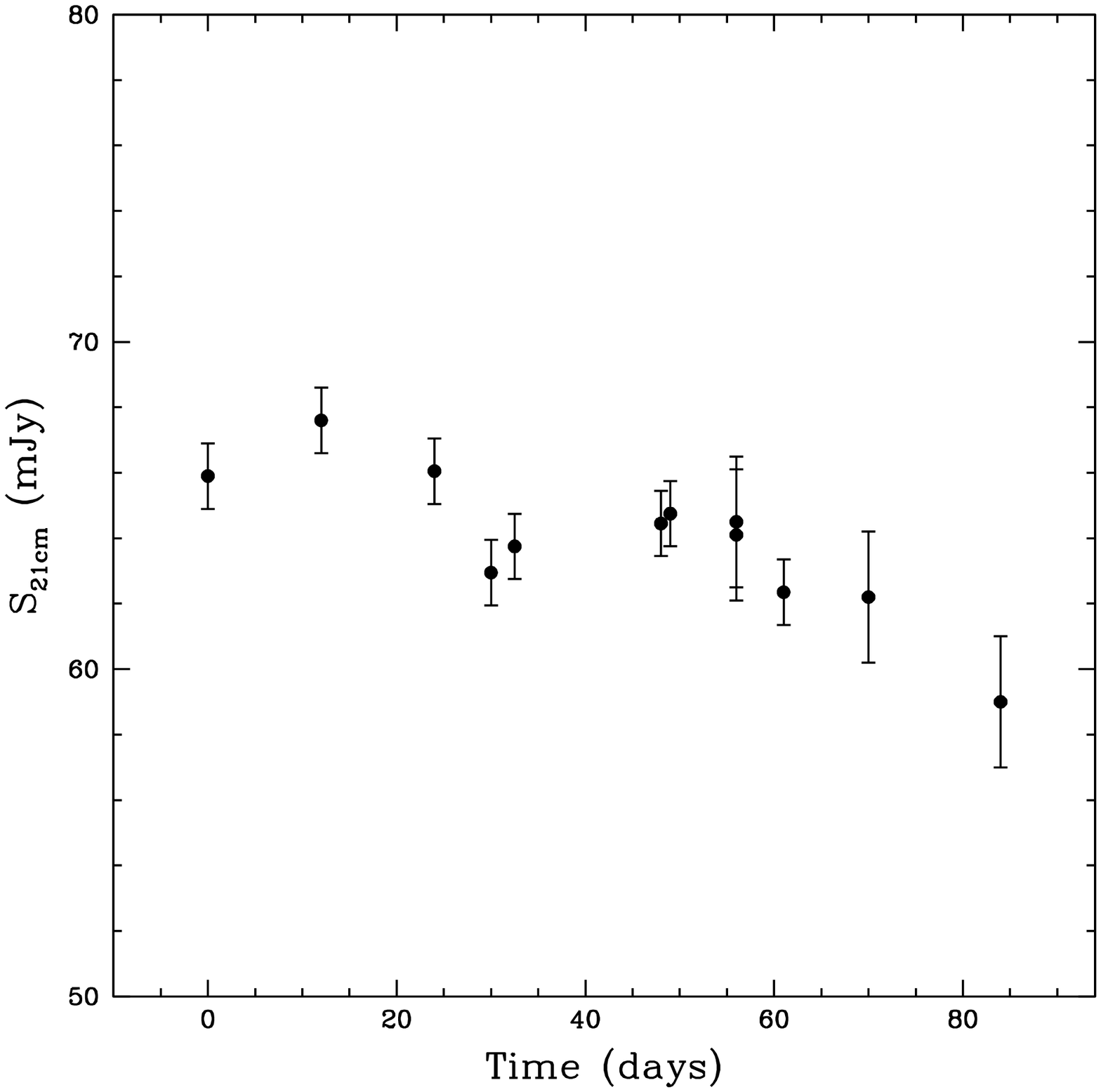}}
\end{center}
\end{figure}

\begin{figure}
\begin{center}
\leavevmode
\vbox{%
\epsfxsize=15cm
\epsffile{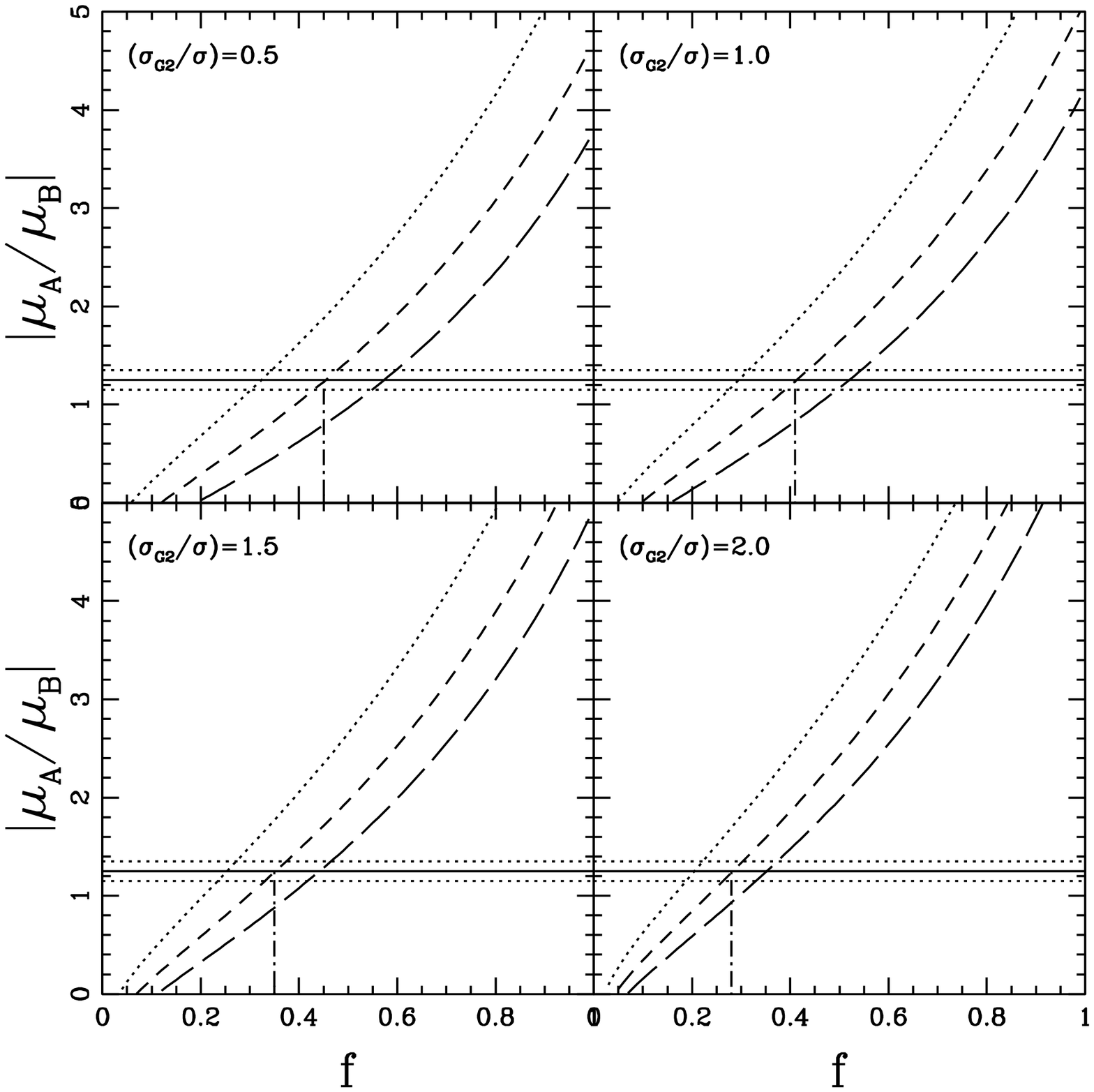}}
\end{center}
\end{figure}

\begin{figure}
\begin{center}
\leavevmode
\vbox{%
\epsfxsize=15cm
\epsffile{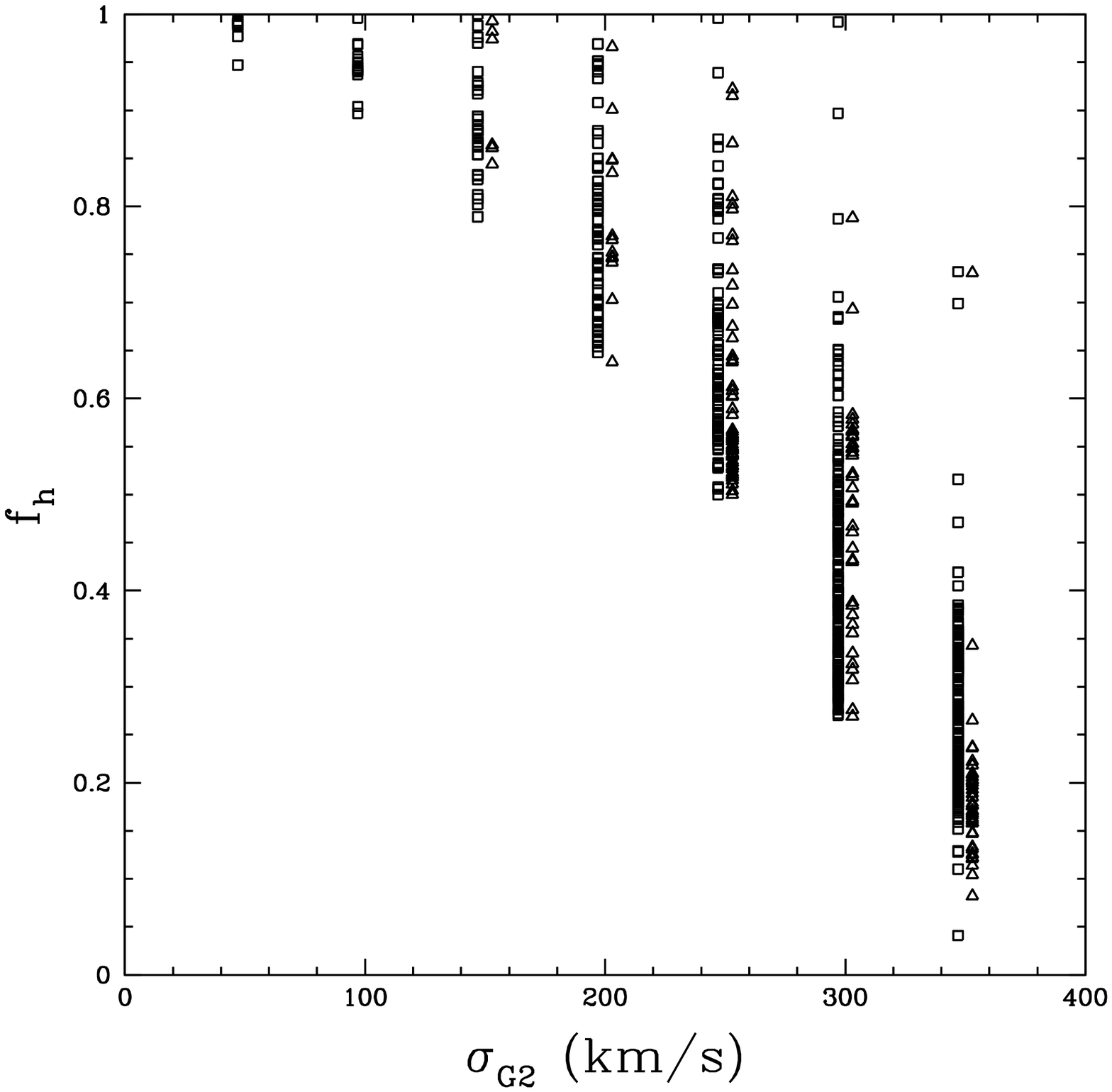}}
\end{center}
\end{figure}

\begin{figure}
\begin{center}
\leavevmode
\vbox{%
\epsfxsize=11cm
\epsffile{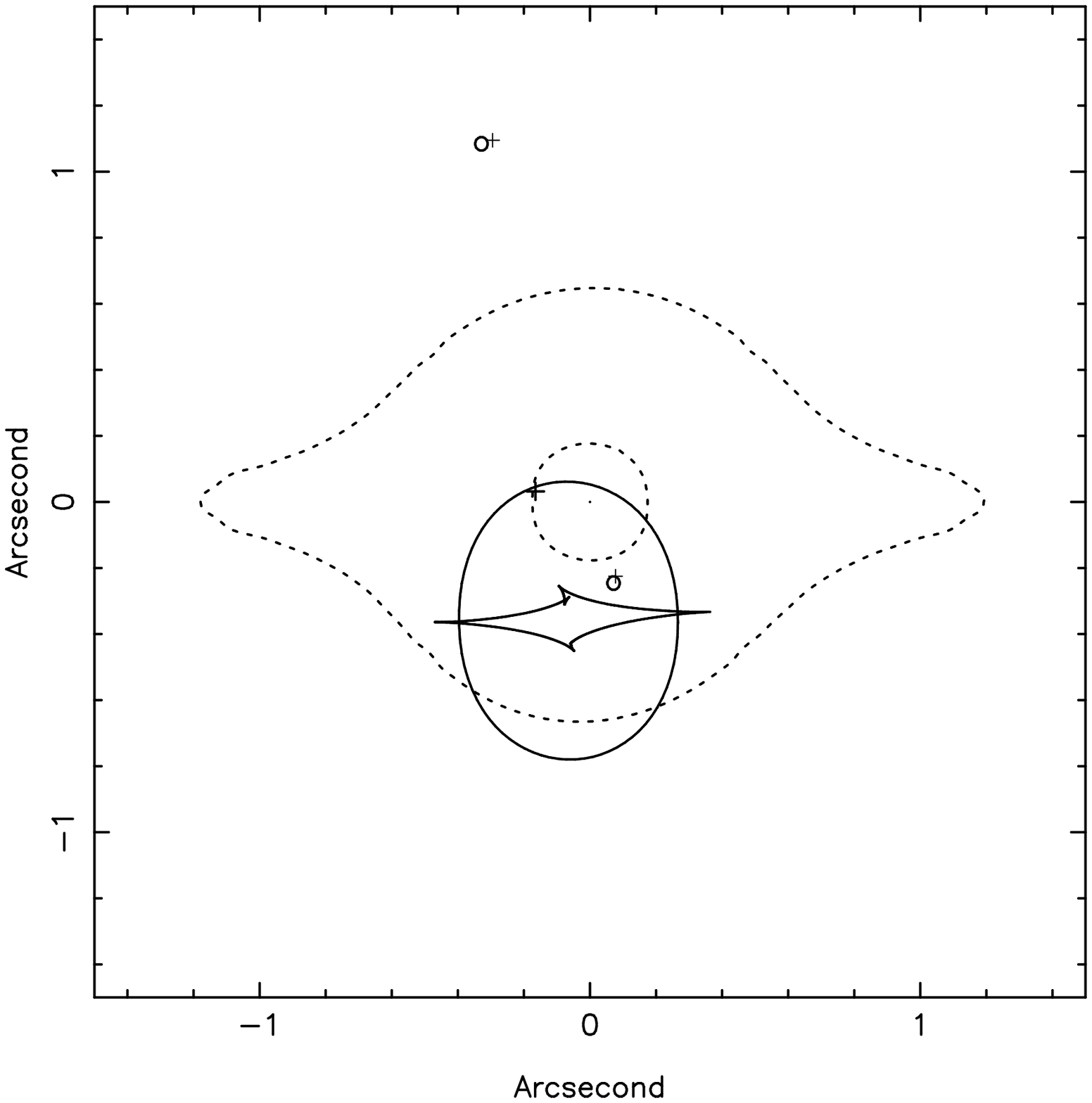}
\epsfxsize=11cm
\epsffile{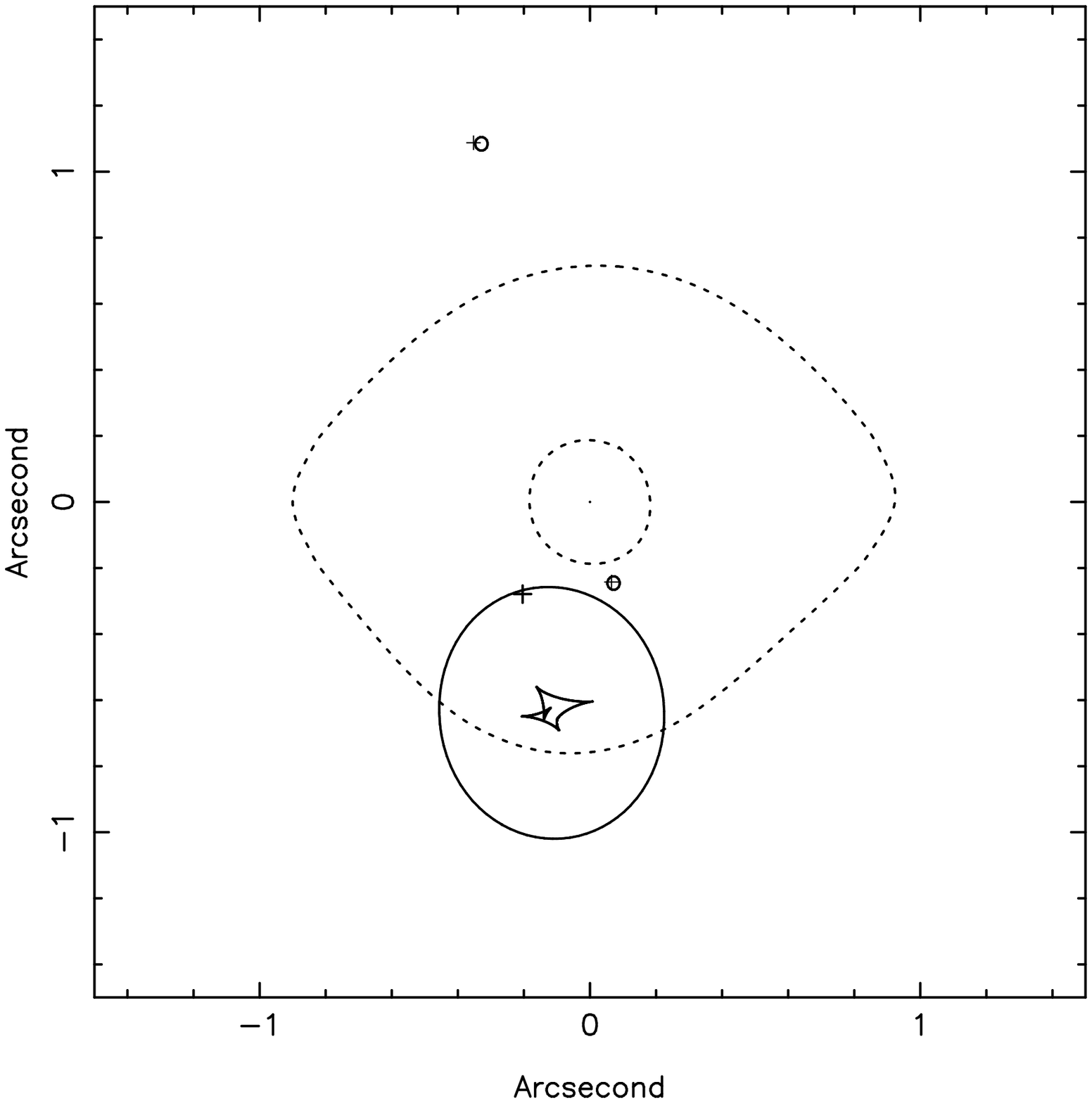}}
\end{center}
\end{figure}

\begin{figure}
\begin{center}
\leavevmode
\vbox{%
\epsfxsize=15cm
\epsffile{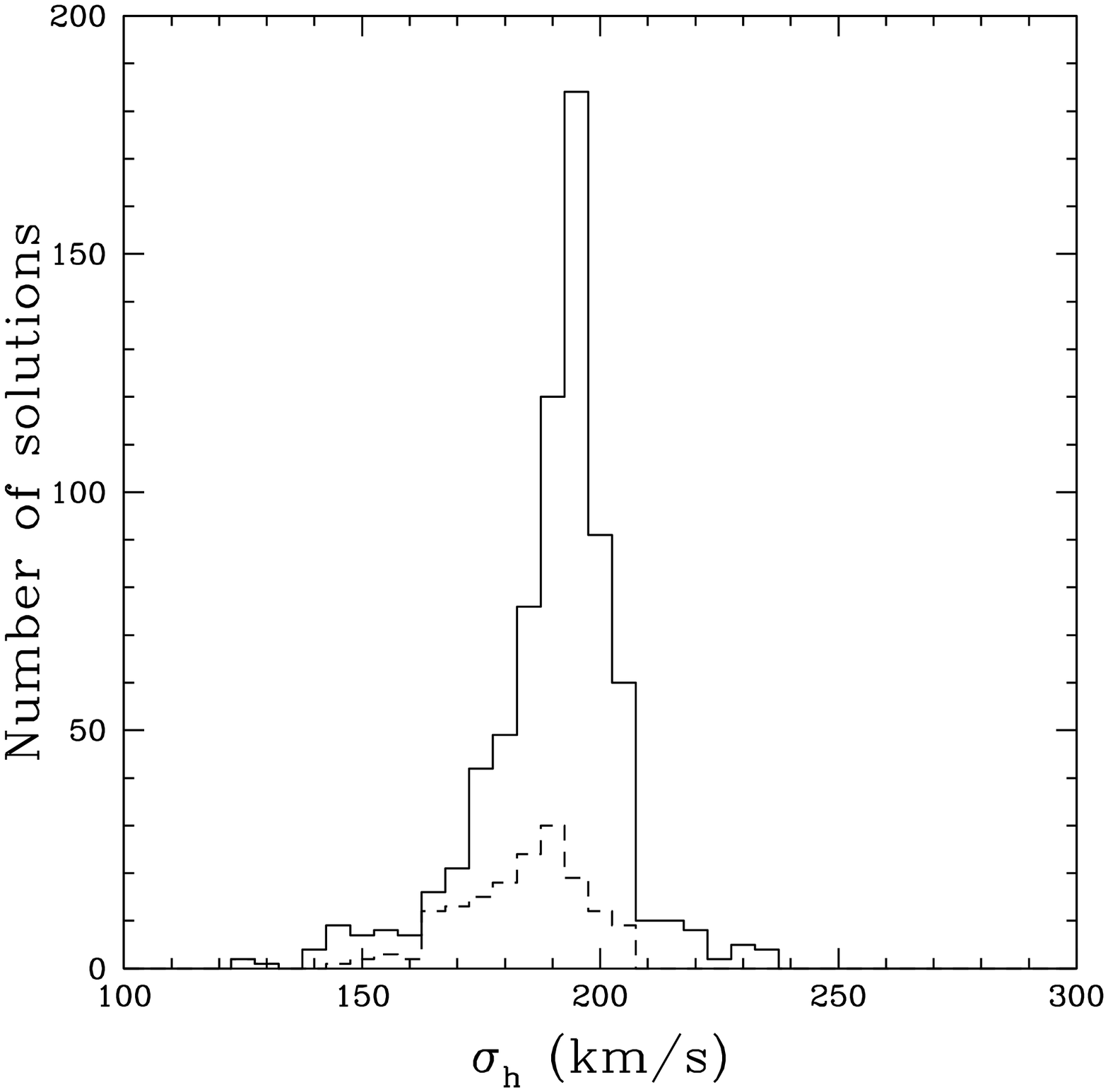}}
\end{center}
\end{figure}

\begin{figure}
\begin{center}
\leavevmode
\vbox{%
\epsfxsize=15cm
\epsffile{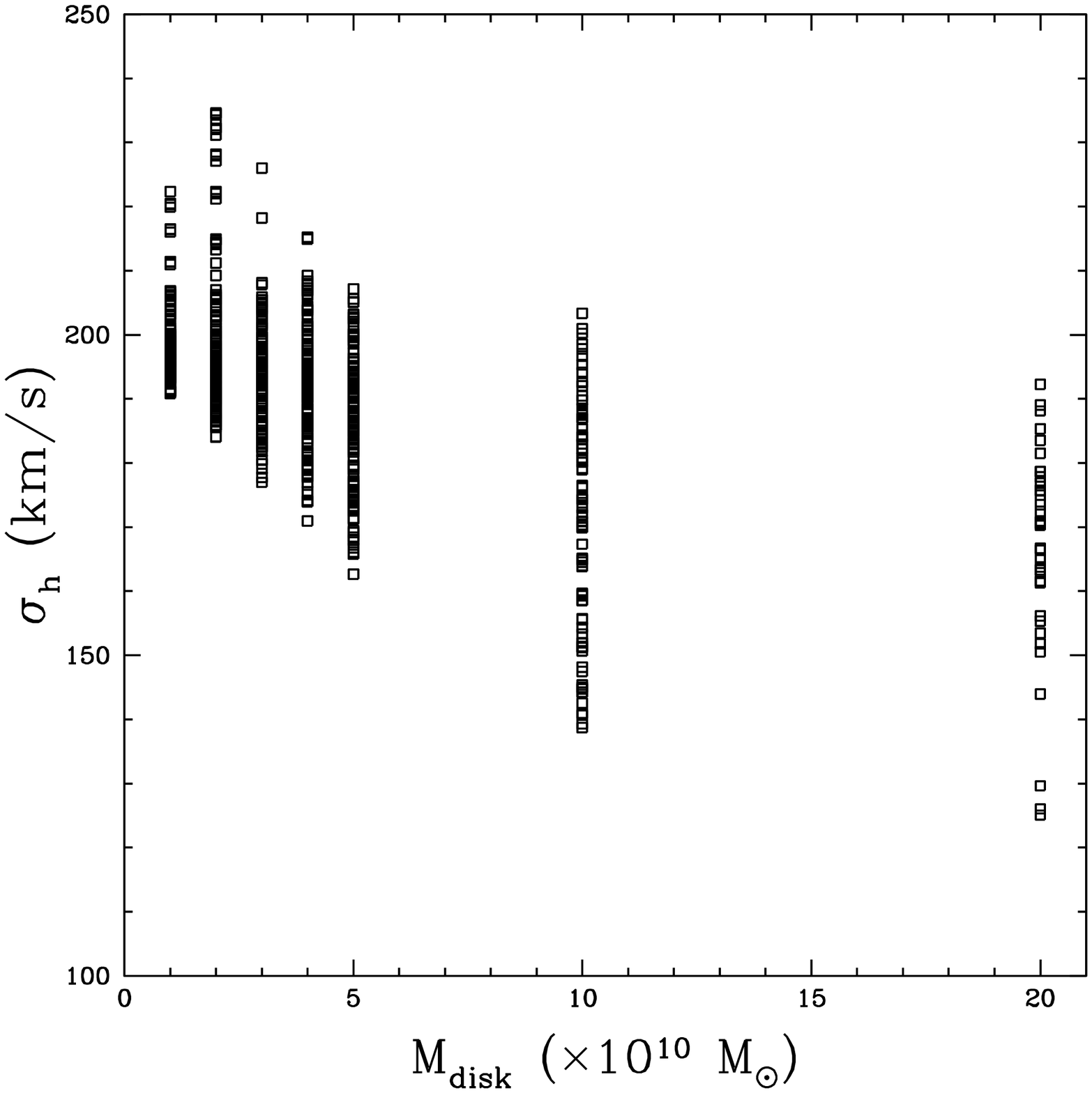}}
\end{center}
\end{figure}

\begin{figure}
\begin{center}
\leavevmode
\vbox{%
\epsfxsize=11cm
\epsffile{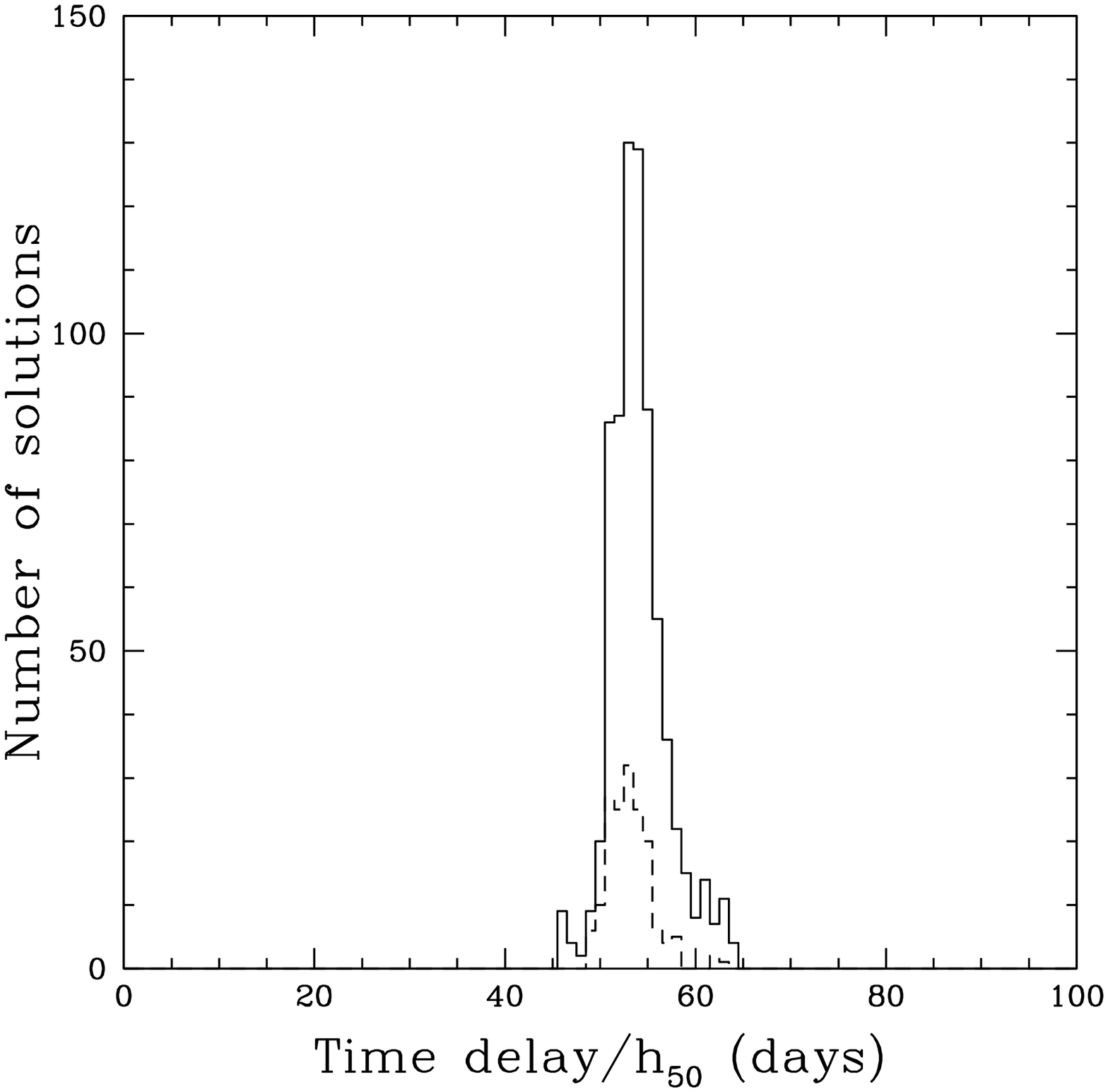}
\epsfxsize=11cm
\epsffile{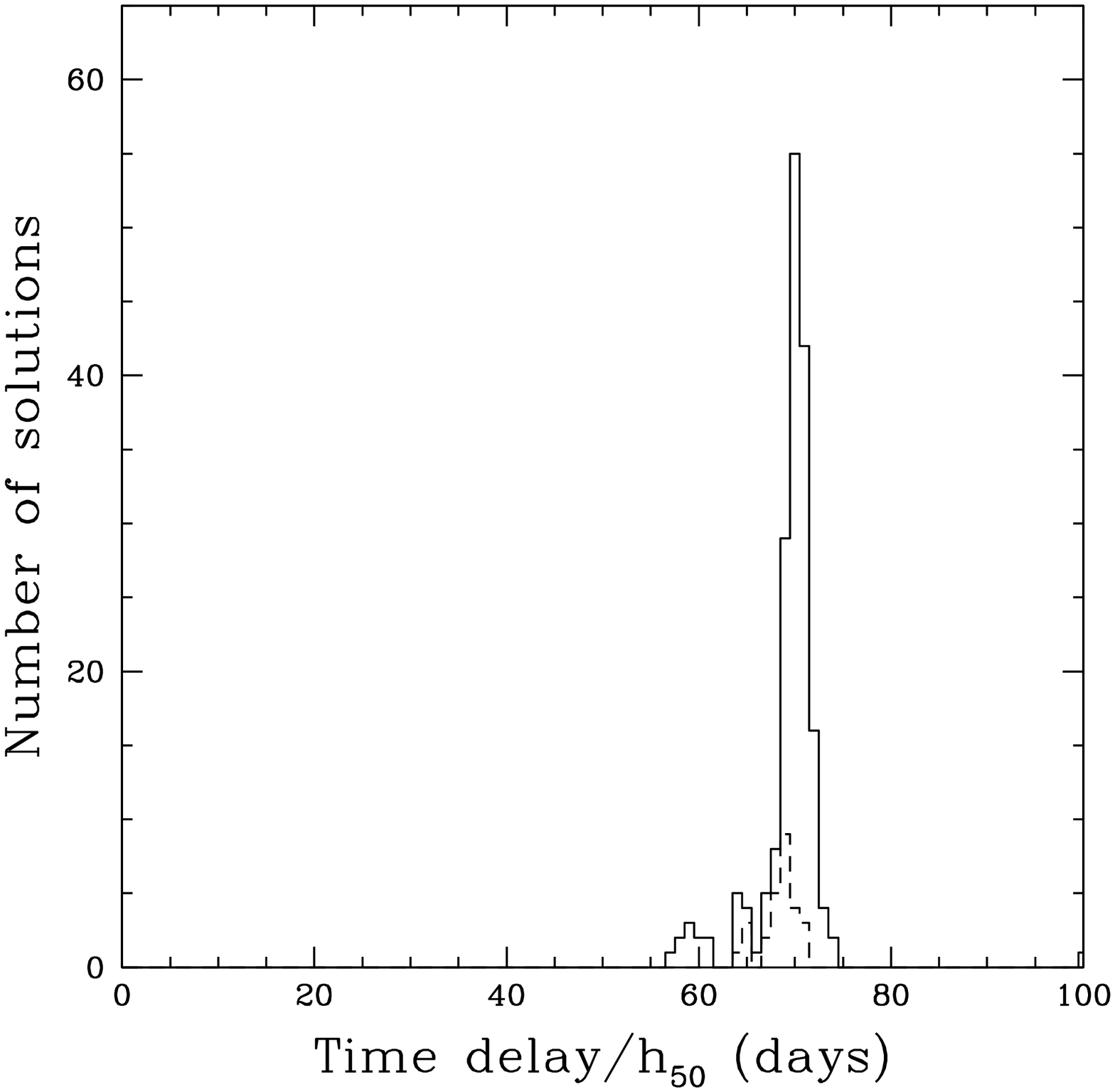}}
\end{center}
\end{figure}


\newpage
\pagestyle{empty}

\begin{table}
\centering
\begin{tabular}{lccccc}
\hline
     &   V (magn.)  &   I (magn.) & $S_{\rm 5}$ & 
                $S^1_{\rm 8.4}$ & $S^2_{\rm 8.4}$\\
\hline
A    &   $22.5\pm0.1$  &  $21.4\pm0.1$ & 45.2 & 58.1 & 28.5 \\
B    &   $23.1\pm0.1$  &  $21.6\pm0.1$ & 37.3 & 48.1 & 23.8 \\
G1   &    ---          &  $20.2\pm0.1$\\
G2   &   $20.6\pm0.1$  &  $19.2\pm0.1$\\
G3   &   $22.5\pm0.1$  &  $21.0\pm0.1$\\
\hline
S1   &   $17.73$       &   --- \\
S2   &   $17.86$       &   --- \\
\hline
\end{tabular}
\caption{Column 1-2: HST V and I band magnitudes of the GL system B1600+434 (November 18, 1995). 
The stars were saturated in I. G1 was not seen in V. Column 3: MERLIN 5 GHz flux density
(mJy) (March 14, 1995). Column 4-5: VLA 8.4 GHz flux densities (mJy) in March 1994 (1) and
August 1995 (2). }
\end{table}

\begin{table}
\centering
\begin{tabular}{lccc}
\hline
          & x$_1$ ($''$) &  x$_2$ ($''$) & Flux \\ 
\hline
A    & -0.33$\pm$0.01 & 1.09$\pm$0.01   &  1.25$\pm$0.10\\
 B    &  0.07$\pm$0.01 & -0.24$\pm$0.01  &  $\equiv$1.00\\
G1   &  0.00$\pm$0.05 & 0.00$\pm$0.05   &\\
G2   & -0.80$\pm$0.10 & -4.40$\pm$0.10  &\\
\hline
\end{tabular}
\caption{Positions and flux ratio of images  and positions 
of galaxies G1 and G2, w.r.t. the defined origin of the coordinate system.}
\end{table}

\begin{table}
\centering
\begin{tabular}{ll}
\hline
Fixed model parameters: &\\
\hline
\hline
 G1: M${}_\rmn{disk}(\times 10^{10}$ M${}_\odot$) & 1.0, 2.0, 3.0, 4.0,\\
        &  5.0, 10.0, 20.0\\
\hline  
 G1: $f_\rmn{disk}$ & 0.1, 0.2, 0.3\\
\hline
 G1: $h_\rmn{disk}$ (kpc) & 1.0, 2.0, 4.0, 8.0, 16.0\\
\hline
 G1: $r_\rmn{c}$ (kpc) & (0.05), (0.10), 0.20, 0.40,\\
        &  0.80, 1.60, [2.40], [3.20]\\
\hline
 G2: $\sigma_\rmn{G2}$ (km/s) & 0, 50, 100, ..., 350\\
\hline
G1: M${}_\rmn{bulge}$ & $\frac{1}{30}\cdot$ M${}_\rmn{disk}$ \\
\hline
G1: $r_{\rm e}$ (kpc) & $\frac{1}{7}\cdot h_\rmn{disk}$ \\
\hline
G1: $f_{\rm bulge}$   & 0.6 \\
\hline
G2: $r_{\rm G2,c}$ (kpc)   & 0.1 \\
\hline
\hline
\end{tabular}
\caption{Parameters used for the NIE and MHP halo models. The parameters within 
parenthesis are only used for the NIE models. The parameters within brackets 
are only used for the MHP models.}
\end{table}

\begin{table}
\centering
\begin{tabular}{lcc}
\hline
           & NIE & MHP\\
\hline
$\mu_\rmn{A}$  &  2.69$\pm$0.26 &  2.03$\pm$0.15 \\
$\mu_\rmn{B}$  & -2.14$\pm$0.22 & -1.64$\pm$0.11 \\
$\mu_\rmn{A}/\mu_\rmn{B}$ &  -1.25$\pm$0.03 &  -1.24$\pm$0.03 \\

\hline
\end{tabular}
\caption{Average magnifications and flux ratios of the source for images A and B.}
\end{table}

\end{document}